\documentclass[10pt,journal,compsoc]{IEEEtran}

\usepackage{comment}
\usepackage{booktabs} 
\usepackage{hyperref}

\usepackage{url}
\usepackage{fancyhdr}
\usepackage{comment}
\usepackage{graphicx}
\usepackage{caption}
\usepackage{multirow}
\usepackage[scientific-notation=true, round-precision=2, round-mode=figures]{siunitx}
\usepackage{array}
\usepackage{color}
\usepackage{amsmath}
\usepackage{amssymb}
\usepackage{mathtools}
\usepackage[ruled,vlined]{algorithm2e}
\usepackage[framemethod=tikz]{mdframed}
\usepackage{subfigure}

\usepackage[shortlabels]{enumitem}
\usepackage{booktabs}
\usepackage{wasysym}
\usepackage{multirow}
\usepackage{threeparttable}
\usepackage[noadjust]{cite}



\begin{document}
%
\title{Improving I/O Performance for Exascale Applications through Online Data Layout Reorganization}

\author{
Lipeng Wan, 
Axel Huebl,
Junmin Gu,
Franz Poeschel,
Ana Gainaru,
Ruonan Wang,
Jieyang Chen,
Xin Liang,
Dmitry Ganyushin,
Todd Munson,
Ian Foster,
Jean-Luc Vay,
Norbert Podhorszki, 
Kesheng Wu,
and Scott Klasky 
\IEEEcompsocitemizethanks{\IEEEcompsocthanksitem L. Wan, A. Gainaru, R. Wang, J. Chen, D. Ganyushin, N. Podhorszki, S. Klasky -- Computer Science and Mathematics Division, Oak Ridge National Laboratory, Oak Ridge, TN 37830. 
E-mail: \{wanl, gainarua, wangr1, chenj3, ganyushindi, pnorbert, klasky\}@ornl.gov
\IEEEcompsocthanksitem A. Huebl, J. Gu, J.-L. Vay, K. Wu -- Lawrence Berkeley National Laboratory, Berkeley, CA 94720. 
E-mail: \{axelhuebl, jgu, jlvay, kwu\}@lbl.gov
\IEEEcompsocthanksitem F. Poeschel -- Center for Advanced Systems Understanding (CASUS), 02826 G\"orlitz, Germany. 
E-mail: f.poeschel@hzdr.de
\IEEEcompsocthanksitem T. Munson, I. Foster -- Argonne National Laboratory, Lemont, IL 60439.\protect\\
E-mail: tmunson@mcs.anl.gov, foster@anl.gov
\IEEEcompsocthanksitem X. Liang -- Missouri University of Science and Technology, Rolla, MO 65409. 
E-mail: xliang@mst.edu
}
}

%
%

\markboth{IEEE Transactions on Parallel and Distributed Systems}
{Data Layout Strategies}
%



\bstctlcite{IEEEexample:BSTcontrol}

\IEEEtitleabstractindextext{%
\begin{abstract}


The applications being developed within the U.S. Exascale Computing Project (ECP) to run on imminent Exascale computers will generate scientific results with unprecedented fidelity and record turn-around time.
Many of these codes are based on particle-mesh methods and use advanced algorithms, especially dynamic load-balancing and mesh-refinement, to achieve high performance on Exascale machines.
Yet, as such algorithms improve parallel application efficiency, they raise new challenges for I/O logic due to their irregular and dynamic data distributions.
Thus, while the enormous data rates of Exascale simulations already challenge existing file system write strategies, the need for efficient read and processing of generated data introduces additional constraints on the data layout strategies that can be used when writing data to secondary storage.
We review these I/O challenges and introduce two online data layout reorganization approaches for achieving good tradeoffs between read and write performance. We demonstrate the benefits of using these two approaches for the ECP particle-in-cell simulation WarpX, which serves as a motif for a large class of important Exascale applications.
We show that by understanding application I/O patterns and carefully designing data layouts we can increase read performance by more than 80\%.





\end{abstract}

\begin{IEEEkeywords}
Parallel IO, Data Layout, IO Performance, WarpX, Data Access Optimization.
\end{IEEEkeywords}}

\maketitle

\IEEEdisplaynontitleabstractindextext

%
\IEEEpeerreviewmaketitle

\IEEEraisesectionheading{\section{Introduction}\label{sec:introduction}}
\IEEEPARstart{E}{xascale} computing has many challenges. One of the most dramatic ones is storing data fast enough. Because storage devices are getting bigger, but not greatly faster~\cite{DiskSpeeds}, exascale systems provide far greater increases in computational speed, relative to petascale computers, than in I/O bandwidth~\cite{IEEESpectrumPerf}. Indeed, the ratio of peak I/O bandwidth to peak compute speed, in bytes per million floating point operations, is getting dramatically worse for current and projected systems~\cite{foster2017computing,CODAR2020}. This growing disparity has profound implications for applications and pushes them toward extremely efficient I/O. Furthermore, many exascale applications are using advanced computing techniques, such as dynamic load balancing and adaptive mesh refinement (AMR), that bring new I/O challenges due to irregular layouts of data in memory.

While applications are moving towards more online analysis and reduction~\cite{foster2017computing}, they still generate a tremendous amount of data for post processing by themselves and other teams.
Thus a key challenge in scientific computing is obtaining near-optimal write performance so that simulations spend minimal time performing I/O. Several projects in the Exascale Computing Project (ECP) focus on methods for optimizing write performance.
For efficiency, these methods generally write data to disk in formats and layouts that mirror those used in memory.
But because these data must subsequently be read for post processing and analysis, 
data should ideally be written to disk in a format and layout that is efficient for both immediate write performance by the simulation and read performance by subsequent analyses. 
However, studies of access patterns~\cite{ 10.1145/1996130.1996139,tian2011edo} have generally considered only  data generated with simple in-memory data arrangements; they have not studied the effect on read performance of more complex memory layouts. Codes that contain multiple data blocks per process permit a wide range of I/O strategies that trade off write performance for potential improvements in read performance. 
These tradeoffs are complicated by the fact that reading can be performed on from one to thousands of nodes, and on each node, we can read from varying numbers of processes and threads.  Each choice can motivate new designs. 

In this paper, we examine these tradeoffs in the context of WarpX~\cite{WarpX2021}, an AMR-based particle-in-cell (PIC) accelerator physics modeling code that uses the AMReX~\cite{AMReX} framework. 
We conduct a comprehensive study of common data layout strategies, focusing in particular on the challenges that WarpX's use of adaptive meshes and its dozens of blocks per process pose for read performance.     
Our results yield insights into when scientists should consider transforming a write-optimized format into a read-optimized format; for example, whether to optimize for reading from the number of nodes from which the data was written, and whether to transform the data after writing but before reading.
WarpX's use of important computational motifs (AMR, dynamic load balancing) make the lessons learned from this study relevant for the larger particle-mesh and AMR communities. 

While we would like to provide general-purpose methodologies that encompass all possible read patterns, we realize that it is not possible to optimize for all. Indeed, the data layout optimization problem is challenging even for some of the simpler read patterns discussed in this paper. Thus we do not address the additional complexity of optimizing for queries, as studied by many groups~\cite{gu2018querying}. Nor do we examine the additional optimizations that are often necessary to obtain good I/O performance on parallel file system such as workload-aware striping~\cite{tian2011edo}, since the GPFS filesystem on Summit supercomputer does not allow applications to configure the striping strategy to optimize the data distribution across storage targets. Such studies are out of scope for this paper. 




Our contributions are as follows. 1) We evaluate the I/O performance of WarpX and quantify performance differences for different data layout strategies. 2) We show that no standard data layout strategy can 
achieve both good write and read performance. 3) We propose and implement an approach to reduce the number of blocks in WarpX data on-the-fly by leveraging the spatial characteristics, which significantly improves the read performance for most common read patterns. 4) By establishing a model based on performance numbers collected from Summit supercomputer, we study the feasibility of using staging techniques to move data asynchronously to a small number of extra staging nodes while
modifying the data layout to optimize for read performance. Our study shows that
not only does this approach improve read performance by up to 85\% compared to the original data layout, but it also can be more efficient in terms of resource utilization compared to a post-hoc data layout reorganization. 


\section{Common Data Layout Strategies}
\label{sec:state-of-the-art}

We describe three data layout strategies commonly used in parallel I/O. 

\subsection{Logically Contiguous Layout}
\label{subsec:contiguous}
In this first strategy, data are linearized and then
stored in their entirety as a logically contiguous block on a file system.
(We say \textit{logically} contiguous because on a parallel file system, the data might be
automatically partitioned into ``stripes'' and distributed across different storage servers) 

\begin{figure}[t] \centering
\includegraphics[width=\columnwidth]{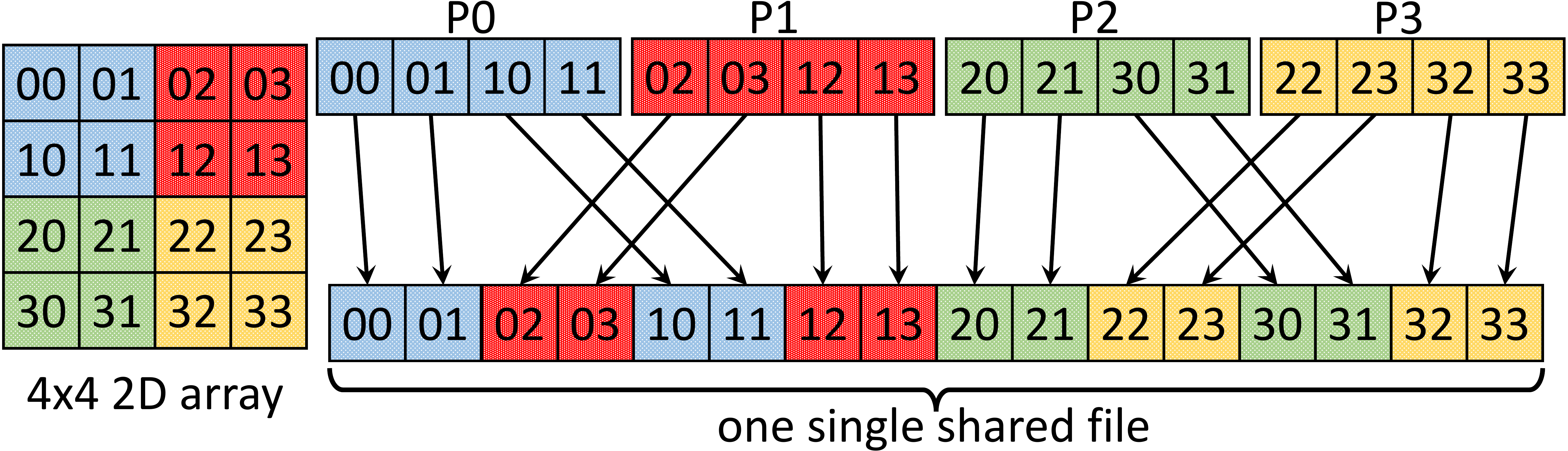}
\caption{Logically contiguous data layout on four processors.
Left: The application's partitioning of a 4$\times$4 2D array.
Above: The data layout in each process' memory.
Below: A logically contiguous layout in the output file.
Arrows show required data movement.
}
\label{fig:contiguous-layout}
\vspace{-3mm}
\end{figure}

Supported by parallel striding I/O operations in the Message Passing Interface (MPI), this intuitive data layout strategy is typically used by default in I/O libraries.
However, it often yields sub-optimal write performance for large MPI-based parallel applications, because while
the partitioned data that resides in each MPI process
may be \textit{locally} contiguous, generating a \textit{globally} contiguous data layout
can require considerable 
inter-process communication. 
For example, 
\autoref{fig:contiguous-layout} shows an MPI program with four processes in which
the data held by each process is linearized in a certain order (e.g., row-major). 
In order to linearize the global
2D array to generate a logically contiguous layout,
data from each process needs to be rearranged. 
Such rearrangements are usually achieved by calling MPI-IO functions, which often trigger some MPI collective operations to coordinate the data movement among processes. 
These rearrangements can be particularly costly for higher dimensional data, especially when many processes participate or when the volumes of the data that needs to be moved are large. 

\subsection{Chunking}
\label{subsec:chunking}
\begin{figure}[ht] \centering
\includegraphics[width=\columnwidth]{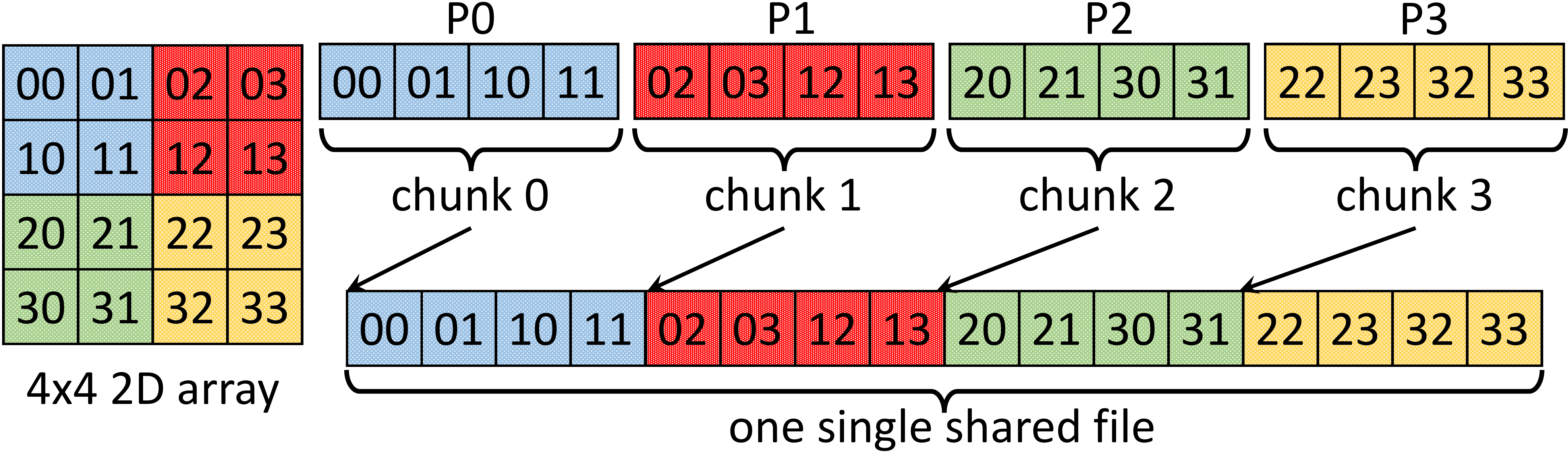}
\caption{Chunked Data layout: 1 chunk per process}
\label{fig:chunked-layout}
\vspace{-1mm}
\end{figure}

Due to the excessive overhead of generating a logically contiguous data layout, a data tiling strategy called \textit{chunking} is often adopted by I/O libraries
as it tends to organize the data produced by large-scale runs of parallel programs in a more efficient manner. The basic idea is straightforward: the original data is not serialized as a single contiguous block but is instead split into multiple chunks that are stored separately in the file, each as a logically contiguous sequence. 
For example, as shown in \autoref{fig:chunked-layout}, each process still operates on a 2$\times$2 block of the global 
array. If we set the
block owned by each process as a chunk, each process is allocated an exclusive space in the file for storing its own data. Thus the overhead caused by rearranging the data among processes is significantly reduced. Moreover, since each chunk can be accessed individually, the performance of operating on a data subset is also improved. 

Parallel HDF5 provides users the flexibility to set chunks with different sizes as long as the chunks are in regular shapes.
However, if the chunk size is too large or the shape of each chunk is not carefully selected, the rearrangement will still be triggered. For instance, in \autoref{fig:chunked-layout}, if each column of the 
global array is set as a chunk, data from two different processes are included in one chunk. In this case, 
the rearrangement 
leads to an increase of inter-process communication. Furthermore, in the general scenarios, the data contributed from individual processes might not form a regular, feasibly large chunk. 
To prevent the rearrangement from slowing down the application's execution, the two-phase I/O techniques~\cite{Tessier2017, 10.1145/3337821.3337875} are often adopted. Usually, such techniques require extra compute nodes/cores dedicated for the data rearrangement.
ADIOS2 does not allow users to set the chunk size based on a logical view of the global space. Instead, it treats each process-local, contiguous data block as a separate chunk.
Moreover, in order to avoid the inter-process coordination for data layout, ADIOS2 adopts a log-structured file format, thus 
the positions of those chunks in the global array is not reflected in the actual data layout. 
Therefore, ADIOS2 requires extra metadata to not only track all the chunks in the actual file, but also record the chunks' positions in the global array. 

\subsection{Sub-filing}
\label{subsec:sub-filing}
\begin{figure}[t]
  \centering
  \subfigure[Without sub-filing]{
    \label{fig:complex-chunking-layout}
    \includegraphics[width=\columnwidth]{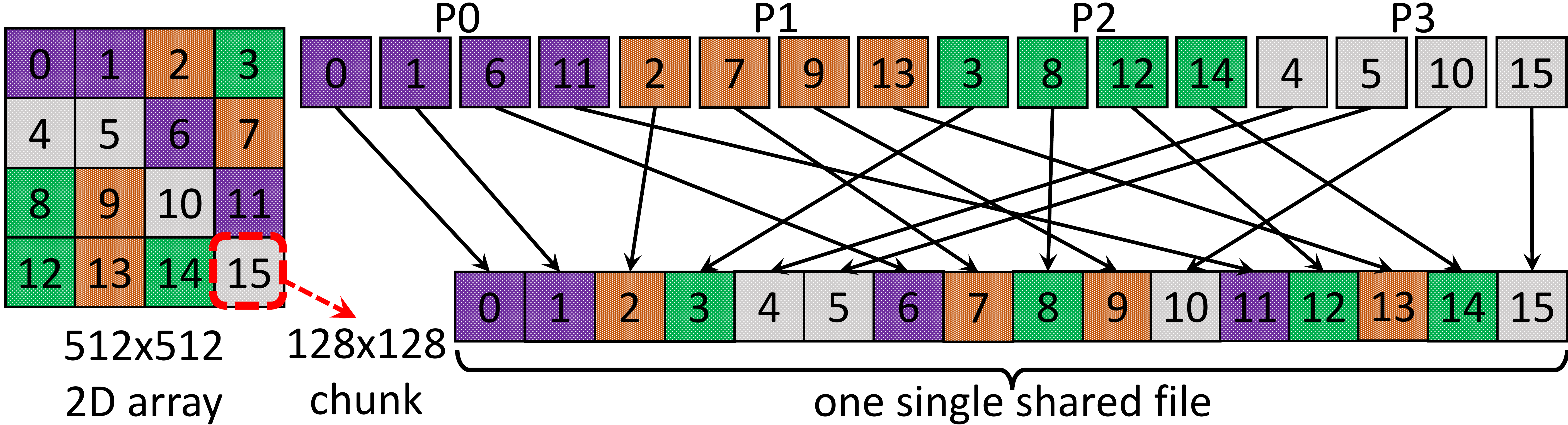}}
  \subfigure[With sub-filing ]{
    \label{fig:complex-chunking-subfiling-layout}
    \includegraphics[width=\columnwidth]{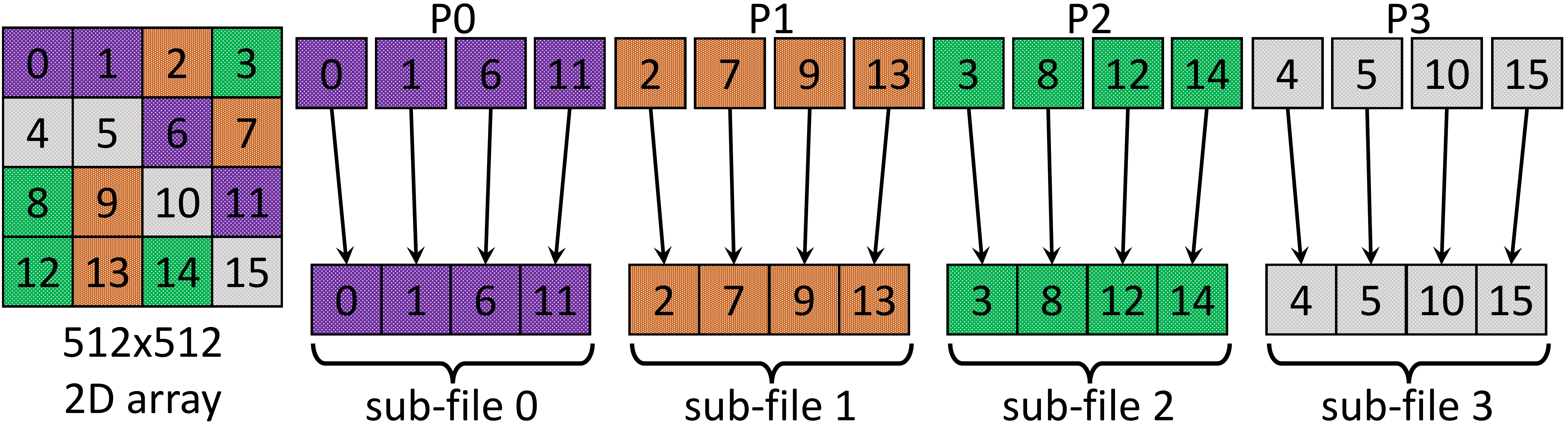}}
    \vspace{-3mm}
  \caption{Data layout: 4 chunks per process, 4 processors}
  \vspace{-5mm}
  \label{fig:with-without-subfiling-diff}
\end{figure}

Although chunking can reduce data rearrangement overhead, the performance of large-scale parallel write operations is sometimes greatly reduced due to file locking contention. 
For example, 
we show
in \autoref{fig:complex-chunking-layout} a configuration in which a 512$\times$512 2D array is distributed over four MPI processes. The four 128$\times$128 chunks on each process
are not always adjacent in the global space. This type of data distribution pattern is often observed in scientific codes that support dynamic load balancing and/or adaptive mesh refinement (AMR). 
Compared to the example illustrated in \autoref{fig:chunked-layout}, where each process only writes one contiguous data chunk to the file, now each process needs to issue four separate write operations, targeting different offsets in the shared file. As a result, file locking contention occurs.

A strategy called \textit{sub-filing} has been developed to address this issue. As shown in \autoref{fig:complex-chunking-subfiling-layout}, instead of writing all data chunks to a single shared file, each process creates an individual file to store its data, to which it writes 
independently to 
take advantage of high parallel file system bandwidth. 
This strategy faces three challenges: 1) Tracking the data chunks stored in each sub-file adds extra overhead to overall I/O performance; 2) Too many sub-files will be created if the parallel program is launched with many processes, overwhelming the parallel file system's metadata service and inconveniencing post-hoc data management; 3) processing of sub-filed data needs to find and open all sub-files that correspond to a read request, which can incur significant read latency. 

Due to these issues, sub-filing is not widely used by I/O libraries as their primary layout strategy. Yet it remains popular as a simple I/O implementation in custom application I/O layers. ADIOS2 uses sub-filing by default; to mitigate the issues caused by sub-filing, it 
provides 
options to aggregate data over a subset of parallel processes through MPI communication, reducing the total number of sub-files, which can lead to significant performance improvements~\cite{Huebl2017}. This approach allows ADIOS2 to trade off a certain degradation in write performance for lighter filesystem metadata load. PnetCDF supports sub-filing, but disables it by default. Parallel HDF5 plans to support sub-filing in the future \cite{HDF5-subfile}.

\section{Data Layout Strategies: A Case Study}
\label{sec:analysis}

Choosing the right data layout strategy is critical for parallel computing programs since it has huge impact on their I/O performance. This is particularly challenging for scientific codes with complex load balancing patterns, often in combination with support for AMR. We use WarpX~\cite{WarpX2021}, an AMReX-based \cite{AMReX} particle-in-cell (PIC) simulation, as an example to study the pros and cons of common data layout strategies for parallel I/O. 

\subsection{Summit's GPFS filesystem and WarpX I/O patterns}
\label{subsec:warpx-overview}
Here we first provide a brief overview of Summit's GPFS filesystem (where we perform all our tests) and WarpX I/O patterns. 
Summit mounts a POSIX-based IBM Spectrum Scale parallel filesystem (Previously it was called GPFS and we use GPFS in this paper for simplicity). This filesystem’s maximum capacity is 250 PB and its maximum performance is about 2.5 TB/s for sequential I/O and 2.2 TB/s for random I/O under FPP mode, which means each process writes its own file~\cite{GPFS}. GPFS, unlike Lustre, does not allow applications to set stripe size and stripe count. The data striping is handled internally in GPFS and not exposed to applications. Therefore, in this study, we only focus on the data layout strategies in user space.

WarpX uses spatial domain-decomposition and dynamic load-balancing to distribute blocked memory structures over MPI processes and associated accelerator devices.
Domain-decomposed data in a physical simulation box consists of meshes (``cells''), which can be of nested refined resolution (MR), and particles.
Between 1 and $N$ blocks of meshes are dynamically assigned to a device, distributed in multiples of a user-defined block size.
Associated with each mesh block is particle data, whose amount can vary arbitrarily in load per mesh block.
Since particles move through the simulation and the operations on particles are more expensive than mesh operations,
WarpX/AMReX provides users with multiple dynamic load-balancing algorithms to achieve a fine-grained, compute-balanced decomposition, which can lead to extremely complex parallel I/O patterns~\cite{Rowan2021}. 

\subsection{Impact of Data Layouts on Write Performance}
\label{subsec:write-performance}

\begin{figure}[ht] \centering
\includegraphics[width=\columnwidth,trim=0 0 0 3.2mm,clip]{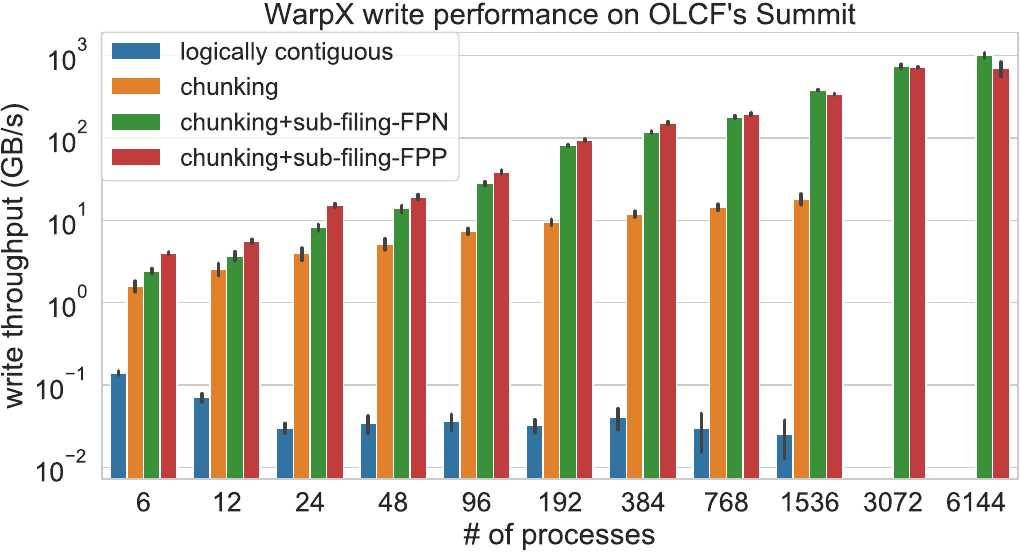}
\vspace{-5mm}
\caption{WarpX write performance on Summit: weak scaling.
(We do not report write performance for ``logically contiguous'' and ``chunking'' strategies with 3072 and 6144 processes because poor write performance did not allow those tests to complete before the time allocation was exhausted.)}
\label{fig:write-performance}
\end{figure}

We set up a weak scaling test on Summit to study how different data layout strategies affect the write performance of AMR-based codes. We use Parallel HDF5 and ADIOS2 to output WarpX simulation data with different layout strategies (using Parallel HDF5 for logically contiguous and chunked data layout while ADIOS2 for the sub-filing strategy), since these two I/O libraries have been integrated into openPMD-api~\cite{openPMDapi}, the I/O framework used by WarpX and other Exascale PIC codes. 
Particularly, we configured the Parallel HDF5 in ``independent I/O'' mode since the ``collective I/O'' mode is not applicable to scientific codes using  load-balancing and/or mesh-refinement due to certain constraints.
In each run, we launch a WarpX simulation with a certain number of compute nodes (up to 1024) with six MPI ranks per node. 
The simulation data is written to the GPFS filesystem using different data layout strategies based on the chosen I/O library. Each compute node writes 64~GB of data per output step and we measure the overall write throughput for all nodes. 
GPFS internally splits big data chunks into 16MB blocks and distributes them across all IO servers. Although the blocks' size cannot be changed in user space, we make sure in our write performance tests the minimal data chunk written by each process is 16MB to take full advantage of the existing setting in GPFS on Summit.
All measurement results are shown in \autoref{fig:write-performance}.


\autoref{fig:write-performance} shows that organizing the WarpX data into a logically contiguous layout significantly degrades write performance. During the execution of the simulation, each process operates on dozens of data blocks for each mesh variable (e.g. 10 blocks per rank in our test). 
While initially adjacent in the global array, these blocks are exchanged among processes as the simulation progresses due to load balancing operations. Thus each process may end up with data blocks from arbitrary locations in the global array. Rearranging these nonadjacent data blocks into a logically contiguous layout can lead to considerable overhead.

Enabling chunking improves the write performance; further significant gains can be achieved by also adding the sub-filing strategy. This is particularly the case when the number of processes increases. As discussed in \autoref{subsec:sub-filing}, the file locking contention becomes a bottleneck if all the processes write all the data chunks to a single shared file concurrently. For instance, when running a simulation with 1,536 processes, 
the overall write throughput of only enabling chunking is still more than one order of magnitude lower than enabling both chunking and sub-filing.

When the sub-filing strategy is enabled, changing the number of sub-files can also affect write performance. As shown in \autoref{fig:write-performance}, when using a small number of processes, creating one sub-file per process (``chunking+sub-filing-FPP'') always achieves higher write throughput compared to one sub-file per compute node (``chunking+sub-filing-FPN''). 
With the increase of processes, the write performance of adopting one sub-file per compute node eventually becomes better. This is because when too many sub-files are created simultaneously, the parallel file system needs to complete many metadata operations, which can slow down the overall write performance.

\subsection{Impact of Data Layouts on Read Performance}
\label{subsec:read-performance}


We tested read performance on Summit using a 2.7~TB WarpX dataset in which each  3D mesh variable is 256~GB.
We focus on these mesh variables, since they were the primary concern for the scientists who ran this type of simulations. 
Because all 3D mesh variables have the same shape and size, we only show read results for one 
variable, ``B''.

\begin{figure}[ht] \centering
\includegraphics[width=\columnwidth]{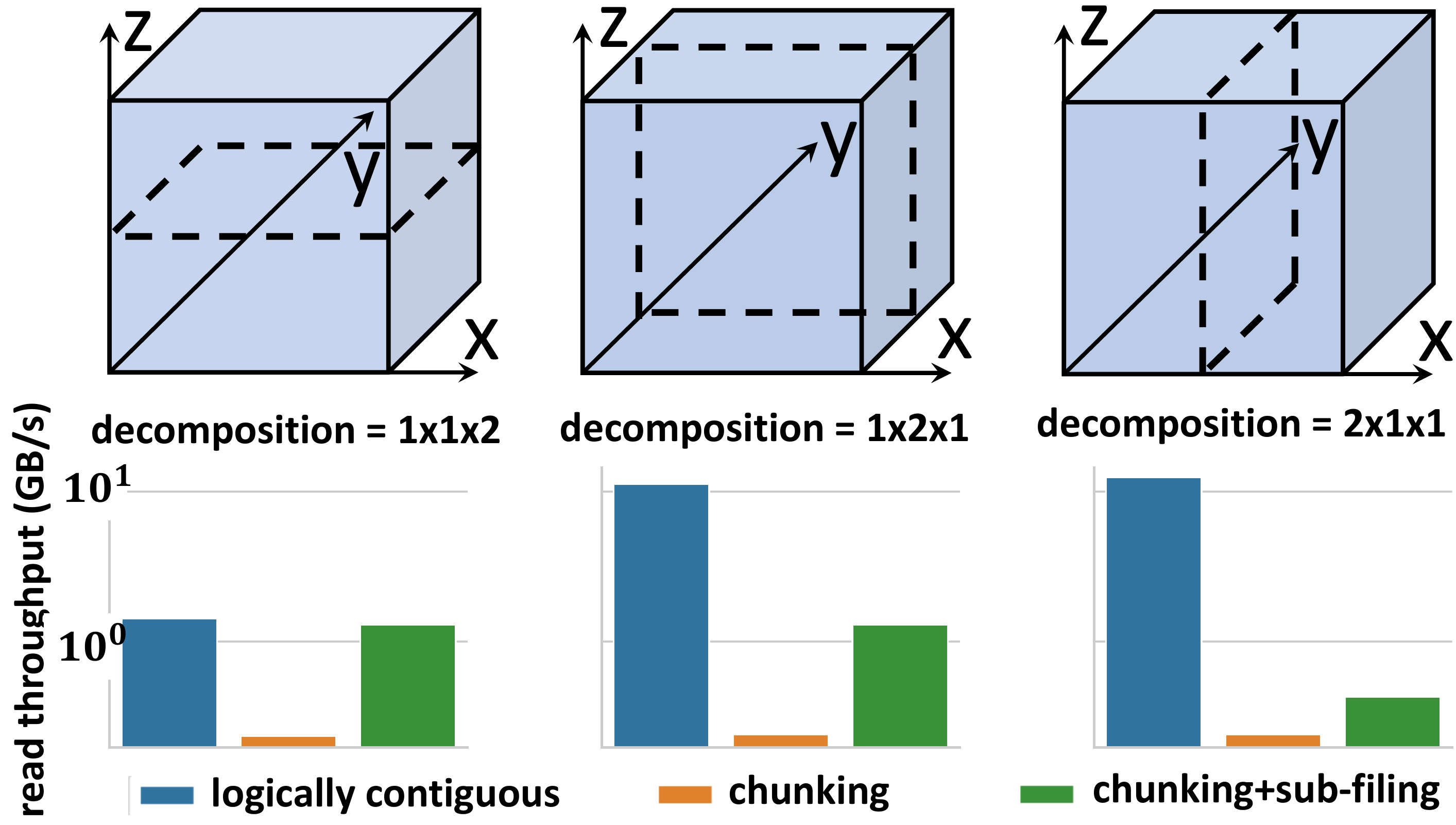}
\caption{Impact of decomposition schemes when reading}
\label{fig:diff-read-decomposition}
\vspace{-4mm}
\end{figure}

\begin{figure}[ht] \centering
\includegraphics[width=\columnwidth]{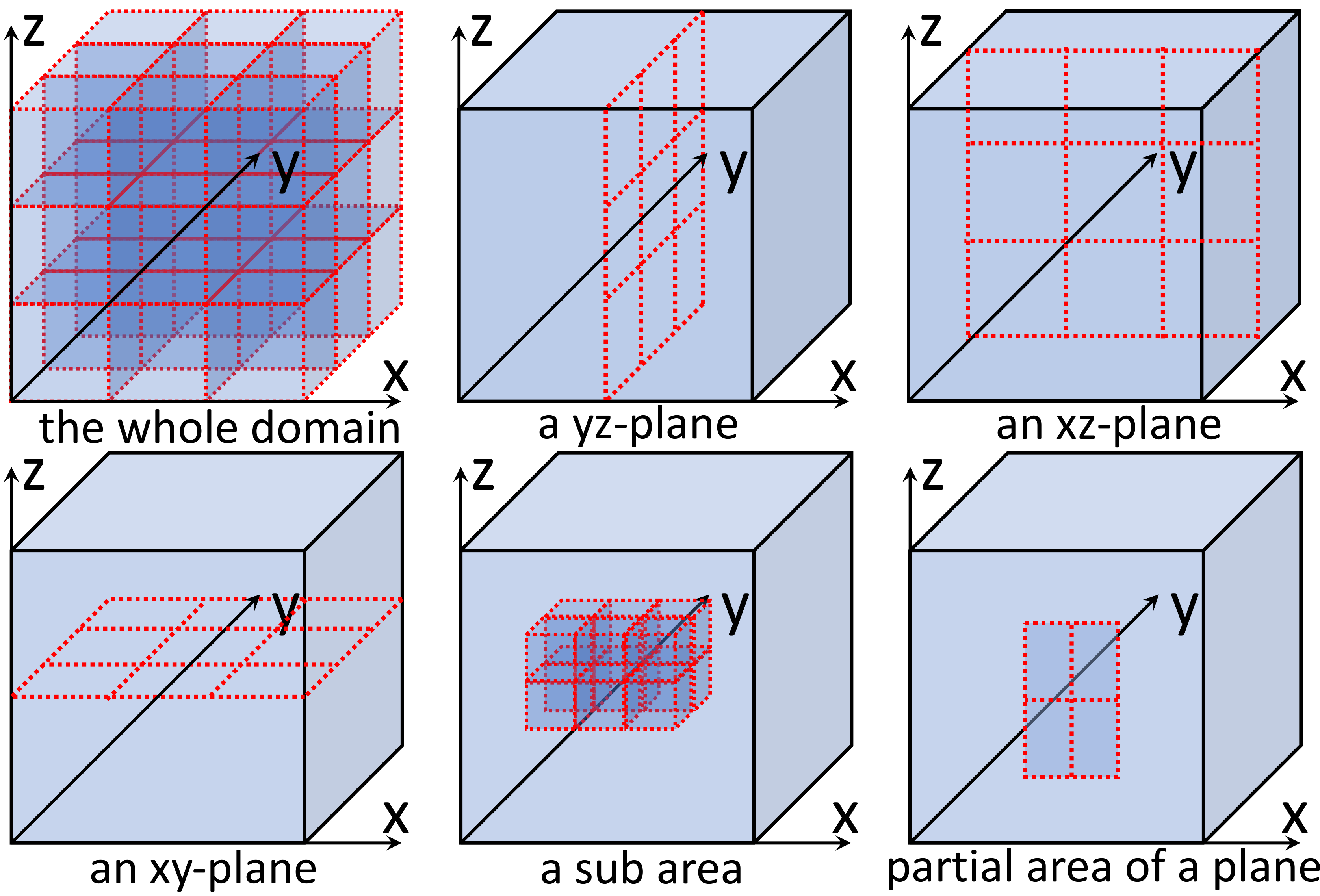}
\caption{Six common read patterns of a 3D mesh variable}
\label{fig:read-patterns}
\end{figure}

To demonstrate how decomposition affects read performance,
we use two Summit nodes 
to read all of ``B''
with three different domain decompositions, with results shown in  \autoref{fig:diff-read-decomposition}.
Decomposition 1$\times$1$\times$2 means the 3D array is split in half along the \emph{z} dimension so that each of the two processes reads half of the data. Similarly, decompositions 1$\times$2$\times$1 and 2$\times$1$\times$1 split the 3D array along the \emph{y} and \emph{x}  dimensions, respectively. 
\autoref{fig:diff-read-decomposition} shows that not only are there large performance differences when reading data generated by different layout strategies under the same decomposition scheme, but read performance is also sensitive to decomposition scheme for certain data layout strategies. 
We will see that this difference has a major consequence when we read
from a larger number of processes. 

\begin{figure*}[ht] \centering
\includegraphics[width=\textwidth]{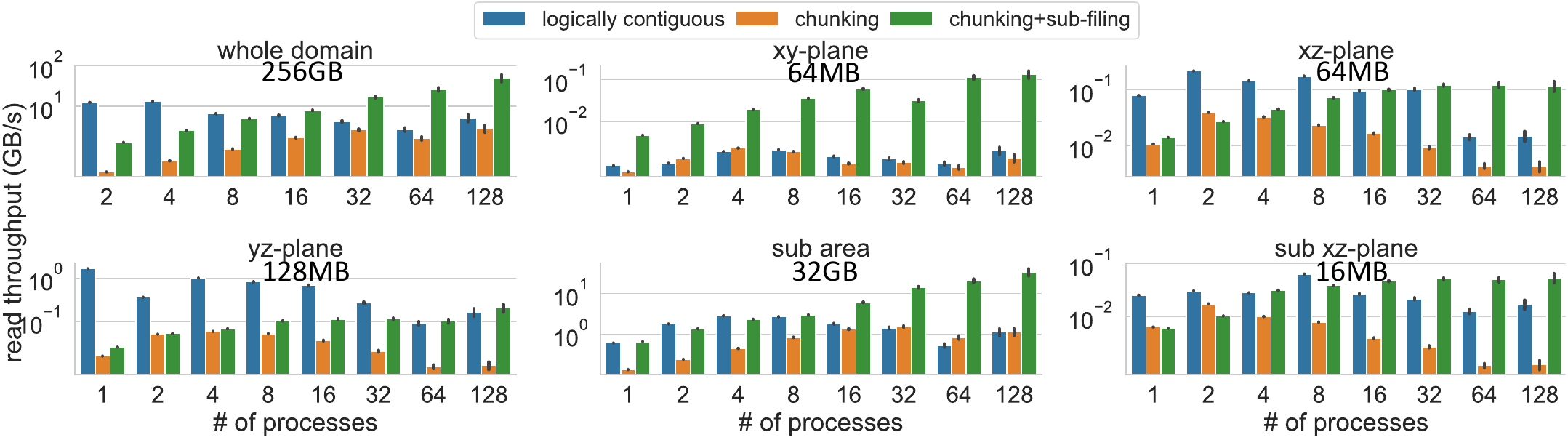}
\vspace{-6mm}
\caption{Performance of reading data generated by common layout strategies under different read patterns}
\label{fig:contiguous-chunking-subfiling-read-perf}
\vspace{-5mm}
\end{figure*}

Finally, we run a comprehensive test with varying number of processes and different decomposition schemes to measure the performance of reading data generated by the common layout strategies under different read patterns.
\autoref{fig:read-patterns} shows six common data access patterns for reading 3D mesh variables in WarpX data. Depending on the number of readers and the decomposition schemes, each reader only reads a portion of the required data concurrently. 
When a number of processes permits multiple possible decomposition schemes, we only present results for the decomposition that achieves the best read performance. 

Our results, presented in \autoref{fig:contiguous-chunking-subfiling-read-perf}, show that the logically contiguous layout offers better read performance in most cases if only a few processes (8 or fewer) are used to read the data.
However, this advantage of logically contiguous layout goes away when reading an xy-plane, as the data elements of an xy-plane are dispersed in a logically contiguous layout of a 3D array.
When the number of processes increases, the performance of reading the logically contiguous layout also decreases. This is because more processes means more concurrent read requests for different regions in the logically contiguous layout, which increases the random seek operations issued to the underlying storage devices. 

When data is generated by only enabling the chunking strategy, read performance is not good in all cases. 
This is because the data read by each process is composed of many chunks; thus, not only must each process find all needed chunks in the file,  it must also linearize those chunks to generate a contiguous array in memory. 
In general, a larger chunk size helps to mitigate this issue, but it also degrades the write performance significantly, as seen in \autoref{subsec:chunking}. 

If both the chunking and sub-filing strategy are enabled when generating data, read performance improves substantially with more processes. We see improvements in read performance compared to the chunking-only case. 
However, reading data with fewer processes is still slow because each process still needs to merge many chunks.

\section{Clustering and Merging of Data Blocks}
\label{sec:merging}

From the performance study presented in the previous section, we observe that none of the common data layout strategies can achieve satisfying write and read performance at the same time. In this section, we study how to leverage the spatial locality in WarpX simulation data to optimize the existing data layout strategies, so that we can find a good tradeoff between the write and read performance.   

\subsection{Spatial Locality}

As mentioned in \autoref{subsec:warpx-overview}, a load balancing operation is often triggered during the execution of AMR-based scientific codes such as WarpX. Once the load balancing operation starts, the data blocks are exchanged among processes. Therefore, as the simulation proceeds, the data blocks each process operates on might be very different from those initially assigned and probably cannot be merged together to form a bigger chunk with a regular shape (e.g., cuboid). This is the fundamental reason why none of the common data layout strategies performs well on WarpX data. For example, if the logically contiguous layout is used, due to the huge number of blocks distributed across all the processes, rearranging the layout will cause a lot of overhead. If the chunking and sub-filing strategy are enabled, although the write performance is improved, reading data from small number of processes becomes slow because each reader needs to read and linearize many small blocks. Parallel HDF5 allows users to set bigger chunks to reduce number of small blocks based on a logical view of the global data space, but in this case that setting hardly matches the physical distribution of data blocks among processes. It is quite possible that data blocks from different processes or nodes belong to the same chunk based on the user setting, resulting in expensive layout rearrangement again.  

\begin{figure}[ht] \centering
\includegraphics[width=\columnwidth]{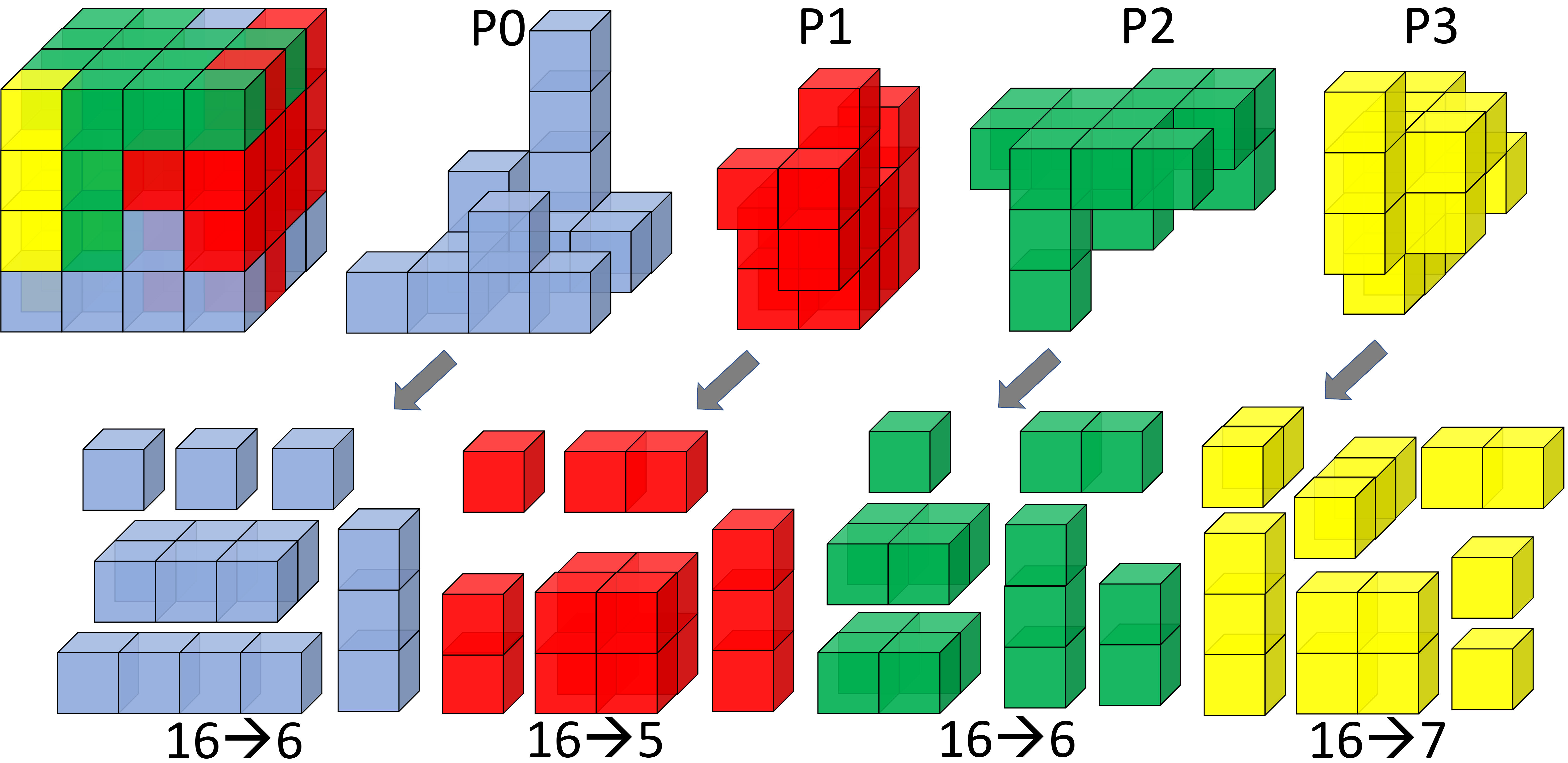}
\vspace{-5mm}
\caption{Merging small blocks into bigger cuboid blocks}
\label{fig:block-clustering-merging}
\vspace{-2mm}
\end{figure}

Our basic idea to address this issue is to leverage the spatial locality in WarpX data to reduce the number of data blocks. Specifically, instead of setting bigger chunks from a logical view of the entire data space, we focus on how data blocks are actually distributed and the possibility of merging data blocks within each process or compute node. If the intra-process or intra-node block merging is possible, we can reduce the number of blocks without causing too much overhead since the intra-process or intra-node data movement is less expensive compared to inter-node data movement. Here is a simple example to demonstrate this idea. As shown in \autoref{fig:block-clustering-merging}, let us assume the mesh variable is a 512$\times$512$\times$512 array which is decomposed into 64 blocks of 128$\times$128$\times$128. After a few rounds of load balancing, the data blocks owned by different processes are in different colors. We see that although it is impossible to merge  \emph{all} data blocks that each process owns, it is possible to merge \textit{some} data blocks into bigger blocks. For instance, some of the blocks on process 0 can be merged, which reduces the number of blocks on process 0 from 16 to 6. Moreover, if process 0 and 1 are running on the same compute node and the data blocks they have can be aggregated together, the number of blocks might be further reduced.    

\subsection{Algorithm for Clustering and Merging Data Blocks}

\begin{figure*}[ht] \centering
\includegraphics[width=\textwidth]{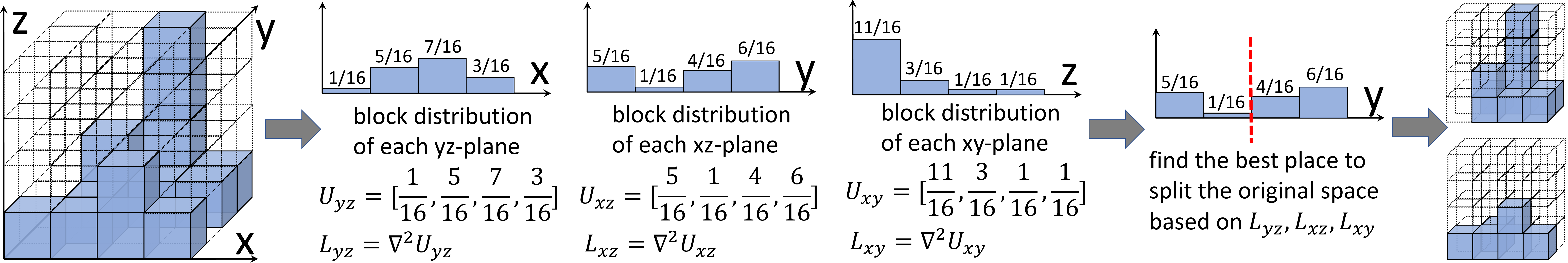}
\vspace{-5mm}
\caption{Finding the best place to split the original space}
\label{fig:block-clustering-algorithm}
\vspace{-2mm}
\end{figure*}

We extend an algorithm proposed by Berger and Rigoutsos \cite{BergerR91} to cluster and merge data blocks in WarpX simulation output. The original algorithm was used to cluster grid points which aims to find minimal number of rectangles to cover all grid points in a 2D space. We have modified the original algorithm so that 1) it can work with 3D data and 2) it will not stop until cuboids filled with original data blocks are found. (In the original algorithm, empty spaces are allowed within each rectangle.)  

\begin{algorithm}[ht]
\footnotesize{
\SetKwInput{KwData}{Input}
\SetKwInput{KwResult}{Output}
\SetAlgoLined
\KwData{Original blocks $[b_0,b_1,\dots,b_{n-1}]$}
\KwResult{Merged blocks $[B_0,B_1,\dots,B_{m-1}]$}
  Find minimal cuboid $C_{root}$ that owns $[b_0,\dots,b_{n-1}]$\;
  Create an empty list $l=[]$ \;
  Add $C_{root}$ to an empty FIFO queue $Q$\;
 \While{$Q$ is not empty}{
  $C_{parent}=Q.dequeue()$\;
  $C_{parent}$ contains original blocks $[b_{p_{0}},\dots,b_{p_{k-1}}]$\;
  \eIf{$Vol(C_{parent}) == \sum_{i=0}^{k-1}Vol(b_{p_{i}})$}{
   $l.insert(C_{parent})$\;
   }{
   Calculate the block distributions\;
   Find the best place to split $C_{parent}$\;
   Split $C_{parent}$ into $C_{left}$ and $C_{right}$\;
   \If{$C_{left}$ has at least one original block}{
    $Q.enqueue(C_{left})$\;
   }
   \If{$C_{right}$ has at least one original block}{
    $Q.enqueue(C_{right})$\;
   }
  }
 }
 \ForEach{$C_i \in l$}
 {
    $C_{i}$ contains original blocks $[b_{i_{0}},\dots,b_{i_{k-1}}]$\;
    Copy $[b_{i_{0}},\dots,b_{i_{k-1}}]$ into memory allocated to $B_i$ 
 }
 \caption{Data Blocks Clustering and Merging}
 \label{alg:block-cluster-merge}
 }
\end{algorithm}
\setlength{\textfloatsep}{0pt}

The input to this algorithm is a set of WarpX data blocks that either are owned by the same process or have been gathered from processes running on the same compute node. For simplicity in presentation, we assume here that these data blocks all have the same shape; in practice, this assumption can be loosened to a certain extent. As shown in Algorithm~\ref{alg:block-cluster-merge}, the first step is to find the minimal cuboid that contains all original data blocks in the global 3D space. For example, the minimal cuboid that contains all data blocks owned by process 0 in \autoref{fig:block-clustering-merging} has shape 512$\times$512$\times$512. Then this cuboid is added to a FIFO queue. If this queue is not empty, a new iteration of the \verb+while+ loop starts. During each iteration of this \verb+while+ loop, a cuboid is first fetched from the FIFO queue, which contains a certain number of the original data blocks. If the volume of this cuboid equals the total volume of all the original data blocks it owns, meaning we have found a bigger block that all the original blocks this cuboid owns can be merged into, we then add this cuboid to a list $l$. Otherwise, there is still empty space in this cuboid and it needs to be further split into smaller cuboids.  

The key component of this algorithm is
finding the best place to split the cuboid.
For example, consider \autoref{fig:block-clustering-algorithm}. To find where to split the cuboid, we need to first calculate the distribution of original data blocks along each dimension. For instance, along the \emph{x} dimension, the cuboid can be divided into four slices of blocks. Within each slice, some positions are occupied by the original data blocks while others are empty. The percentages of original data blocks within each slice form a vector $U_{yz}=[\frac{1}{16},\frac{5}{16},\frac{7}{16},\frac{3}{16}]$. This vector represents a histogram that can be treated as a 1D binary image. Finding the best place to split on the \emph{x} dimension is similar to detecting the edge in this image. Therefore, we can apply common edge detection technique such as the Laplacian edge detector to this histogram. For instance, $L_{yz}=\nabla^{2}U_{yz}$ is the result of Laplace’s differential operator applied to $U_{yz}$. To find the location of the edge in this histogram, we need to find a zero-crossing in $L_{yz}$; as the Laplace operator is a second order derivative, a zero-crossing corresponds to an inflection point in $U_{yz}$, which is the edge location we are looking for. Similarly, we can calculate $L_{xz}$ and $L_{xy}$ for the \emph{y} and \emph{z} dimensions, respectively, and find their zero-crossings also. Among all these zero-crossings, we select the one with the steepest slope in the histogram as the place to split the entire cuboid. In this example, that place is on the \emph{y} dimension.   

After the current cuboid is split into two sub-cuboids, we check whether these two sub-cuboids are empty or not. If they still own original data blocks, we add them to the FIFO queue for further processing. Once the FIFO queue becomes empty, the while loop exits and all the cuboids filled up with original data blocks are in list $l$, meaning the clustering of data blocks is finished. Now we need to merge the actual data of these original data blocks. Specifically, for each cuboid in list $l$, we merge and serialize the data of all the original blocks it owns into a bigger memory buffer as a new data block. All these new data blocks will then be written out as the ``chunks'' to the parallel file system.
 
\subsection{Evaluation on Data Blocks Clustering and Merging}

\begin{figure*}[ht] \centering
\includegraphics[width=\textwidth]{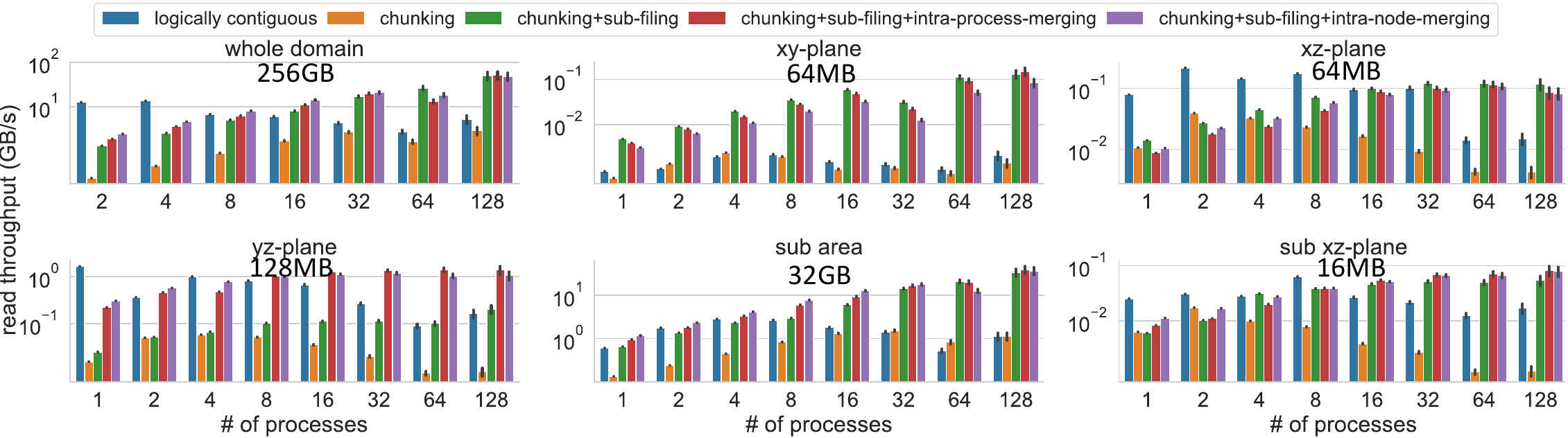}
\vspace{-5mm}
\caption{Read performance when intra-process or intra-node blocks merging is enabled}
\label{fig:contiguous-chunking-subfiling-intra-process-merge-intra-node-merge-read-perf}
\vspace{-2mm}
\end{figure*}

\begin{figure}[ht]
  \centering
  \subfigure[Total time saved using intra-process block merging]{
    \label{fig:intra-process-time-saved}
    \includegraphics[width=\columnwidth,trim=0 0 0 8mm,clip]{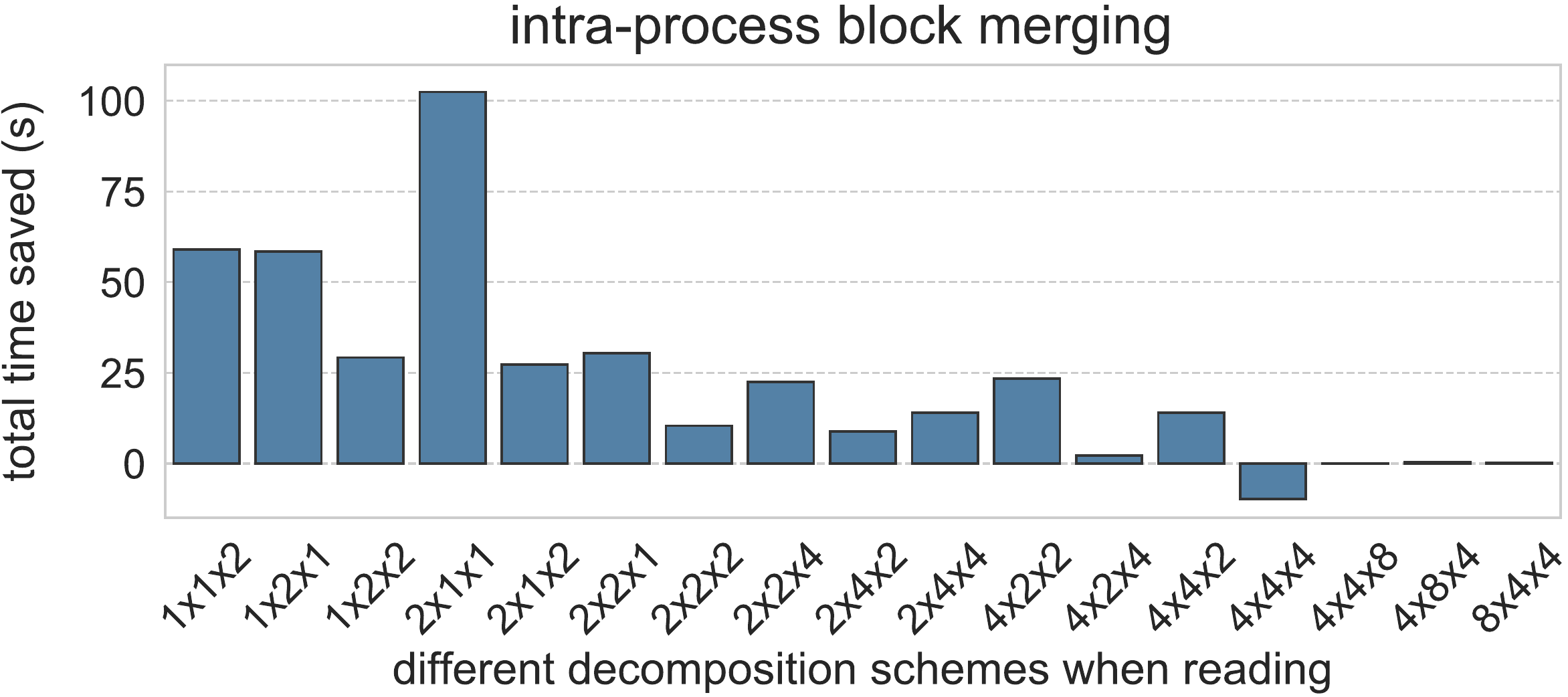}}
  \subfigure[Total time saved using intra-node block merging]{
    \label{fig:intra-node-time-saved}
    \includegraphics[width=\columnwidth,trim=0 0 0 8mm,clip]{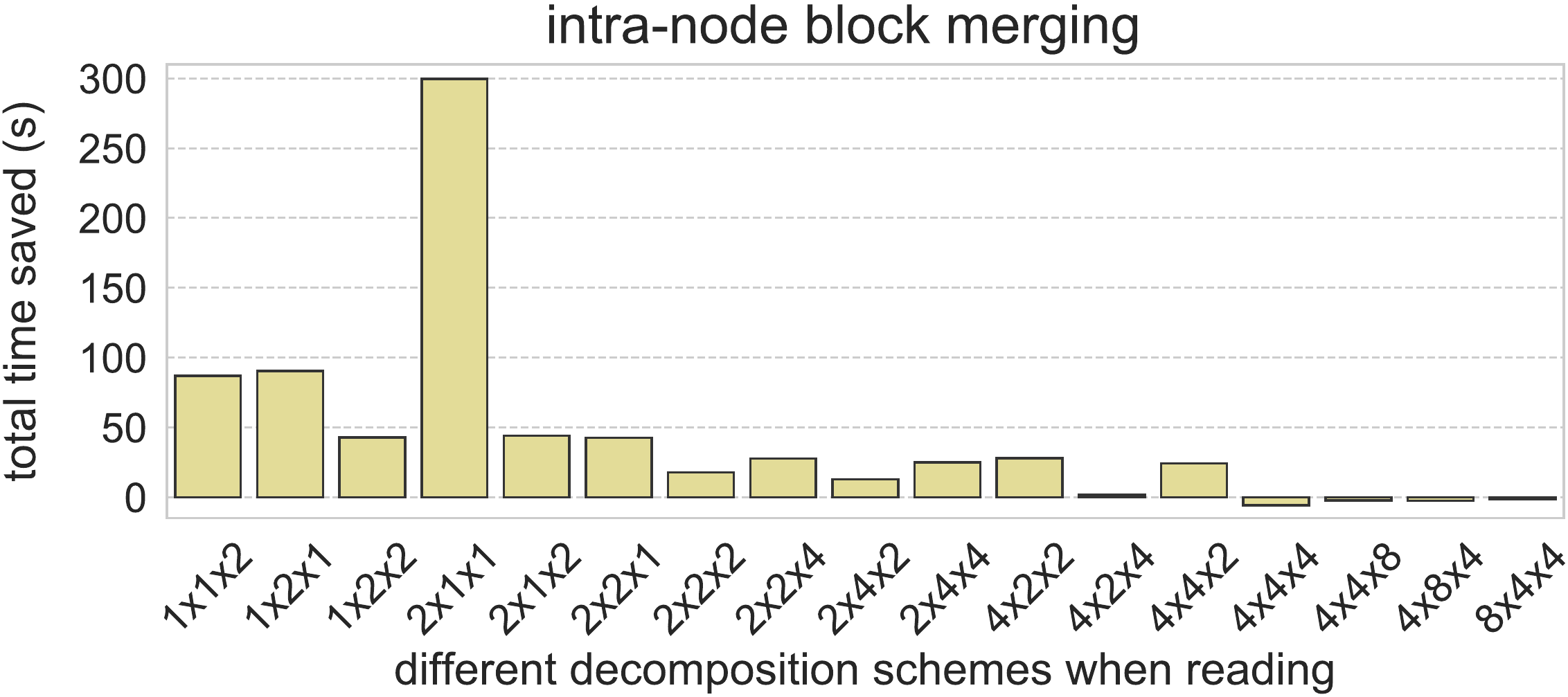}}
    \vspace{-3mm}
  \caption{Writing and reading a 3D mesh variable: total end-to-end time saved. Comparison for intra-process and intra-node block merging.}
  \label{fig:reading-whole}
\end{figure}

\begin{figure}[ht]
  \centering
  \subfigure[Node-second gain and loss using intra-process block merging]{
    \label{fig:intra-process-gain-loss}
    \includegraphics[width=\columnwidth,trim=0 0 0 8mm,clip]{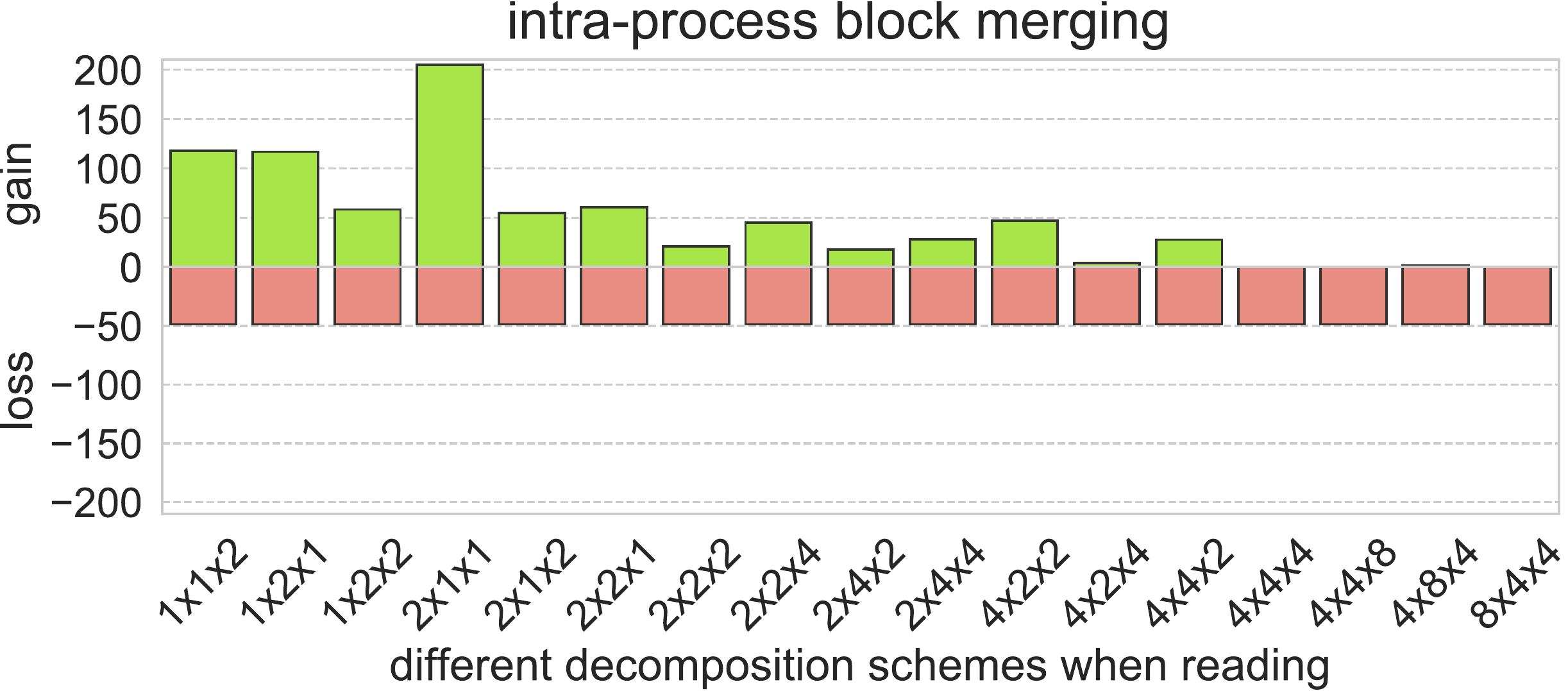}}
  \subfigure[Node-second gain and loss using intra-node block merging]{
    \label{fig:intra-node-gain-loss}
    \includegraphics[width=\columnwidth,trim=0 0 0 8mm,clip]{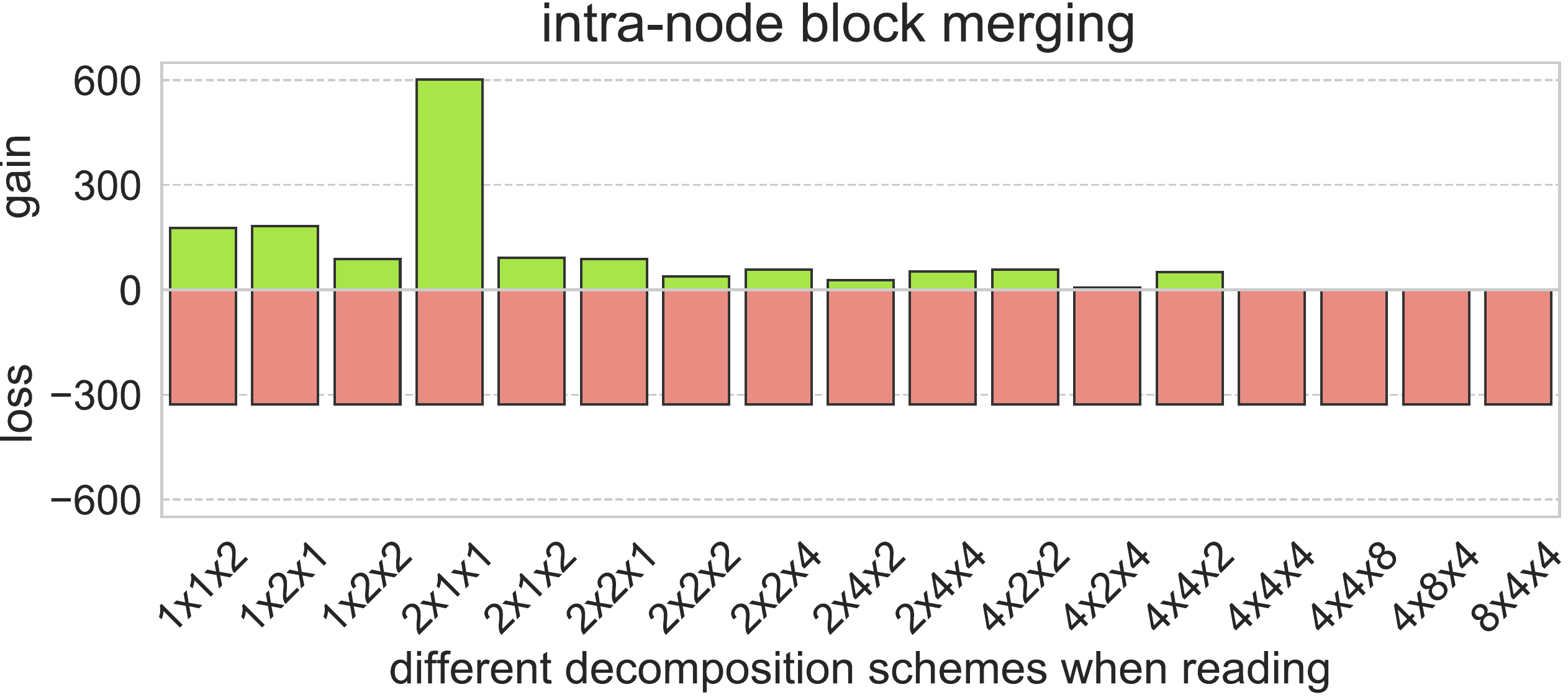}}
     \vspace{-3mm}
  \caption{Writing and reading a 3D mesh variable: node-seconds saved on the reader side vs. node-seconds lost on the writer side.}
  \label{fig:reading-whole}
\end{figure}

We evaluate the data block clustering and merging algorithm from two aspects: 1) the read performance improvements it provides, and 2) the write overhead it causes. Particularly, we apply the algorithm to the real WarpX simulation output. For the WarpX run with 256 compute nodes and 1,536 processes, each process operates on roughly 10 original data blocks and there are about 64 data blocks on each compute node. If the intra-process blocks clustering and merging are enabled, on average the number of data blocks each process operates on is reduced from 10 to 3. If the intra-node blocks clustering and merging are enabled, on average the number of data blocks on each compute node is reduced from 64 to 10. Intra-node clustering and merging require data blocks owned by processes running on the same compute node to be gathered to one process, leading to extra data movement overhead. 

By reducing the number of data blocks, the read performance is expected to be improved. As shown in \autoref{fig:contiguous-chunking-subfiling-intra-process-merge-intra-node-merge-read-perf}, we measure the performance of reading the WarpX data after intra-process and intra-node blocks clustering and merging respectively, and compare the performance numbers with those achieved by using common data layout strategies. As we can see, under four of the six common read patterns (reading whole domain, yz-plane, sub area, xz-plane), the read performance of enabling data blocks clustering and merging is notably better than only enabling chunking and sub-filing. However, since the merged data blocks might be different from each other in terms of shape and size (e.g., as shown in \autoref{fig:block-clustering-merging}), enabling block merging cannot guarantee read performance improvements under every read pattern (e.g., when reading the xz-plane and xy-plane). 

We also evaluate the overall benefit of enabling the block clustering and merging by taking both the performance improvements and overhead into account.
When the intra-process blocks clustering and merging are adopted, the overhead of clustering blocks of one 3D mesh variable in WarpX data is less than 0.001 seconds, while merging these data blocks in memory takes about 0.19 seconds. If we enable intra-node block clustering and merging, these two numbers are 0.0003 and 1.03 seconds, but there is an extra overhead of about 0.25 seconds in gathering data blocks within each compute node through an MPI collective operation. 
On the reader side, the seconds saved by enabling the block merging is obtained by calculating the difference between read times without and with block merging. Therefore, the total time saved by using block merging is calculated by subtracting the overhead caused on the writer side from the time saved on the reader side. As shown in \autoref{fig:intra-process-time-saved} and \autoref{fig:intra-node-time-saved}, for most scenarios, enabling block merging shortens the total time of the application workflow.

If scientists are more sensitive to the cost of a job running on supercomputers (which is usually measured by the product of the compute nodes and time that the job occupies), here we also calculate the node seconds that we gain and loss by turning on the clustering and merging of data blocks to assess its usefulness.
As shown in \autoref{fig:intra-process-gain-loss}, when intra-process block merging is enabled, the loss in node seconds per variable on the writer side is always 256$\times$(0.001+0.19)=48.9. The gain in node seconds is then the number of nodes used for reading multiplied by the number of seconds saved by enabling the block merging. We see in the figure that if the number of readers is less than four, the gain of node seconds in the reader is always greater than the loss in the writer, indicating that we should enable the intra-process block merging. By enabling intra-process block merging, we can save up to 150 node seconds for writing and reading each 3D mesh variable. However, with the increase of readers, the gain of node seconds on the reader side drops, which makes intra-process block merging less efficient. We did the same calculations for the intra-node block merging case. As shown in \autoref{fig:intra-node-gain-loss}, since intra-node block merging causes more overhead and the writer side usually uses more compute nodes, only when the decomposition scheme 2$\times$1$\times$1 is used for reading, the gain of node seconds is greater than the loss. 
One thing we must point out is that the outputs of large-scale long-running simulation jobs are usually repeatedly accessed by domain scientists after being generated. Therefore, even if the block clustering and merging incurs some overhead when generating the data, there will often be a long-term benefit in downstream data workflows.

In summary, both intra-process and intra-node block merging can improve the read performance, but intra-process block merging achieves better resource utilization for scientific codes with dynamical load-balancing and/or AMR on supercomputers. For common scientific campaigns where \textit{data is written once but read many times}, these approaches can offer significant benefits in the long run.

\section{Reorganization of Data Layout}

Although the data blocks clustering and merging approach proposed in  \autoref{sec:merging} improves the read performance compared to only enabling the chunking and sub-filing strategies, 
the lack of flexibility in adjusting the size and shape of the merged blocks prevents it from achieving even better read performance.
In this section, we study the feasibility of reorganizing the data layout on-the-fly 
to further improve the read performance if extra system resources are available.

\begin{figure}[bht]
  \centering
  \subfigure[Post-hoc data layout reorganization]{
    \label{fig:posthoc}
    \includegraphics[width=\columnwidth]{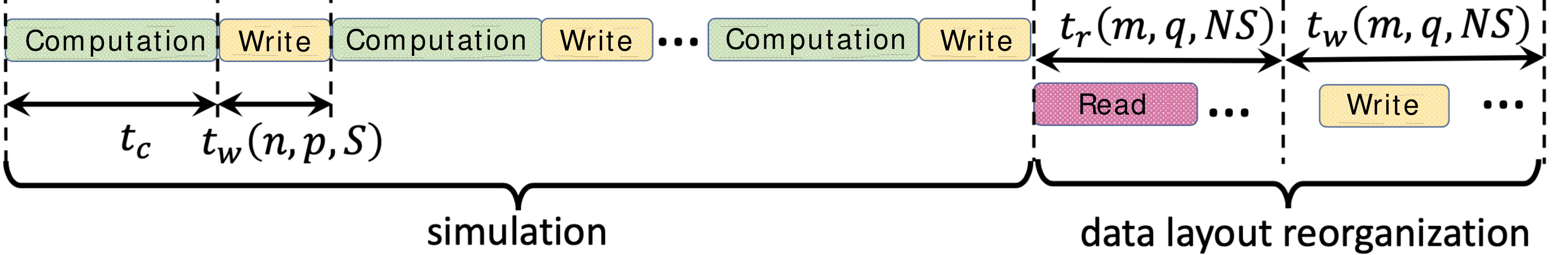}}
  \subfigure[On-the-fly data layout reorganization]{
    \label{fig:on-the-fly}
    \includegraphics[width=\columnwidth]{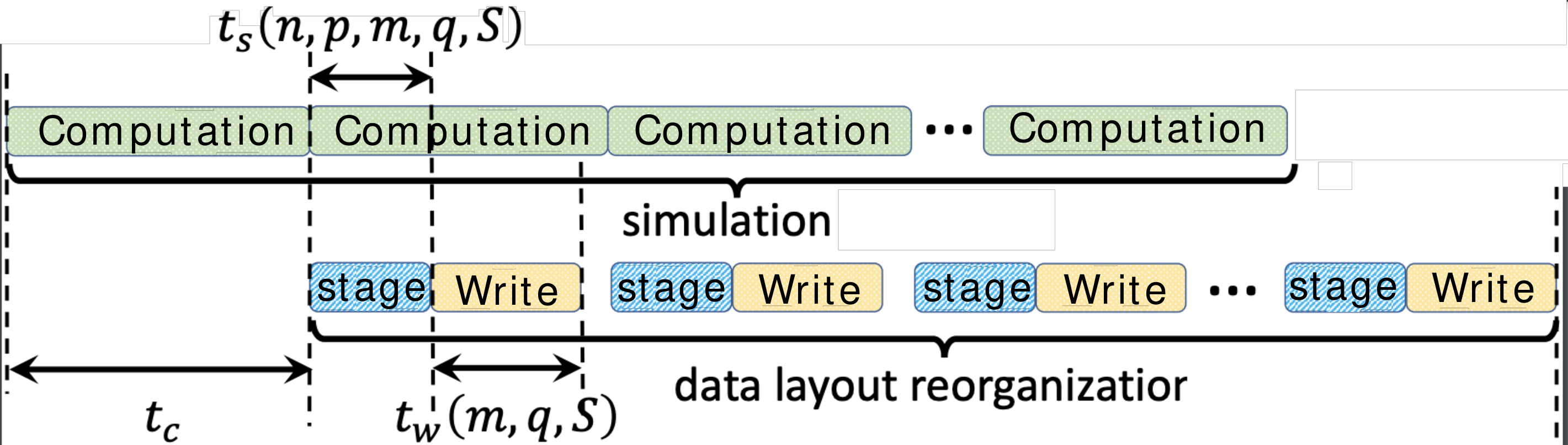}}
    \vspace{-2mm}
  \caption{Data layout reorganization}
     \vspace{-6mm}
  \label{fig:otf-vs-post}
\end{figure}

\subsection{Different Ways to Reorganize The Data Layout}

As shown in \autoref{fig:posthoc}, let us assume we run a simulation on Summit which periodically outputs data with chunking and sub-filing enabled to achieve the optimal write performance. The most naive way to reorganize the data layout is to launch a job with certain number of processes after the simulation finishes, which reads in the data generated by the simulation and writes it out again to the parallel file system with a new layout. This is the so-called post-hoc data layout reorganization. The advantage of this approach is that it does not slow down the simulation job. Its disadvantage is domain scientists cannot access the data in reorganized layout immediately after the simulation job finishes, because reading in and writing out the data might take a long time, especially when the simulation output is large.

An alternative approach is to reorganize the data layout on-the-fly. Specifically, while the simulation is running, instead of letting the simulation write the outputs directly to the parallel file system, we move the data to certain processes through MPI or other communication protocols to form the data layout we need, and then let those processes write the data to the file system. 
Existing studies have proposed to implement this functionality based on different techniques, including two-phase I/O and staging. For example, Tessier et al.~\cite{Tessier2017} developed an efficient topology-aware two-phase I/O algorithm to aggregate contiguous pieces of data before performing reads/writes.
Kumar et al.~\cite{10.1145/3337821.3337875} proposed a two-phase approach that leverages MPI-based data aggregation to reorganize particle data layout on the fly. Solutions based on the staging techniques were also proposed in~\cite{10.1145/1551609.1551618,10.1145/2110205.2110209,10.5555/2388996.2389022}. Although these approaches showed promising results for certain application use cases, we still need to understand if it is feasible or efficient to reorganize complex data layouts on the fly for scientific applications that use dynamic load-balancing and/or AMR since moving the huge amount of data blocks among processes can cause significant overhead.

\subsection{Efficiency of Data Layout Reorganization}
In order to study this, here we leverage the non-blocking staging technique called Strong Staging Coupler~\cite{SSC} to asynchronously move the data blocks from the compute nodes where the simulation is running, to a few staging nodes where the data blocks are merged as a contiguous chunk before being written to the parallel file system. We choose this technique instead of others because it is part of ADIOS2 library which can be called by WarpX through the openPMD-api~\cite{openPMDapi}.
We run tests on Summit to collect performance numbers and then build a model which is quantified based on these performance numbers. We list the symbols used in this model in \autoref{tab:notations-of-symbols}. 

\begin{table}[ht]
  \caption{Symbols used in our model}
  \centering
  \scalebox{0.95}{
    \begin{tabular}{|c|l|}
      \hline
      \textbf{Symbol} &  \textbf{Notation} \\
      \hline
      $t_c$ & Computation time between two outputs\\ 
      \hline
       $t_{w}()$& Time to write data to the PFS\\ 
      \hline
       $t_{r}()$& Time it takes to read data from the PFS\\
      \hline
       $t_{s}()$& Time it takes to stage data\\
      \hline
      $n$ & \# of compute nodes used for simulation \\
      \hline
      $p$ & \# of processes per node used for simulation \\
      \hline
      $m$ & \# of compute nodes used for data layout reorganization \\
      \hline
      $q$ & \# of processes per node used for data layout reorganization \\
      \hline
      $S$ & Size of each simulation output \\
      \hline
      $N$ & Total \# of simulation outputs \\
      \hline
      $U$ & Resource utilization in compute node seconds \\
      \hline
    \end{tabular}
  }
  \vspace{-4mm}
  \label{tab:notations-of-symbols}
\end{table}

\begin{figure}[ht] \centering
\includegraphics[width=\columnwidth]{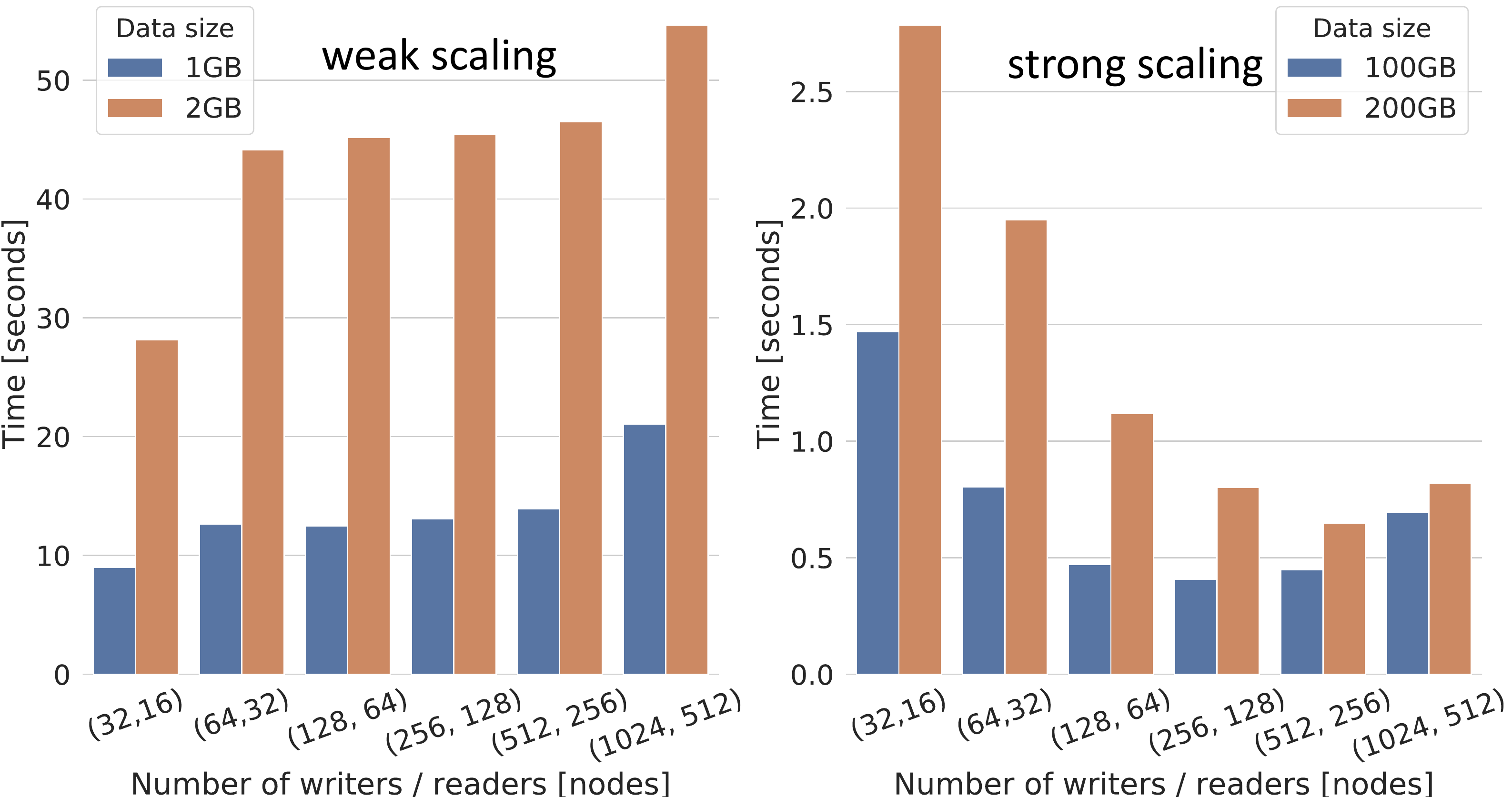}
\caption{Weak and strong scalability tests for staging}
\label{fig:staging-scale-test}
\vspace{-2mm}
\end{figure}

Among all these symbols, $t_{w}()$, $t_{r}()$ and $t_{s}()$ are dependent on the scale and parallelism of the setup, which means they are functions of number of compute nodes, number of processes per node and size of data. Thus, we first present results of some weak and strong scalability tests for the staging technique we used to understand how $t_{s}()$ ($t_{w}()$ and $t_{r}()$ have been studied in previous sections) changes when different setups are adopted. As shown in \autoref{fig:staging-scale-test}, in our weak scaling test, we fix the data size per writer node (1 or 2~GB) and measure the data staging time when different number of writers and readers are used. In our strong scaling test, we fix the total size of the data (100 or 200~GB) and also measure the data staging time when different number of writers and readers are used. From the results we can see that the data staging time $t_{s}()$ cannot be characterized by a simple formula, which adds extra complexities to this problem.

Now let us introduce our model. As shown in \autoref{fig:posthoc}, the total simulation time is $N[t_c+t_{w}(n,p,S)]$, while the time it takes to reorganize the data layout offline is $t_{r}(m,q,NS)+t_{w}(m,q,NS)$. Therefore, the resource utilization of post-hoc data layout reorganization is $U=nN(t_c+t_{w}(n,p,S))+m[t_{r}(m,q,NS)+t_{w}(m,q,NS)]$. For on-the-fly data layout reorganization, there are two possible scenarios: 1) If $t_{s}(n,p,m,q,S)+t_{w}(m,q,S) \leq t_c$, then the execution time of the entire workflow is $Nt_{c}+t_{s}(n,p,m,q,S)+t_{w}(m,q,S)$, thus the resource utilization can be calculated as $(n+m)[Nt_{c}+t_{s}(n,p,m,q,S)+t_{w}(m,q,S)]$. 2) If $t_{s}(n,p,m,q,S)+t_{w}(m,q,S) > t_c$, the computation will be delayed and not continue until the current simulation output is written to the file system. In that case, the execution time of the entire workflow is $t_c+N[t_{s}(n,p,m,q,S)+t_{w}(m,q,S)]$ and the resource utilization is $U=(n+m)\{t_c+N[t_{s}(n,p,m,q,S)+t_{w}(m,q,S)]\}$. For a given simulation setup, some variables in these formulas are fixed, including $n,p,S$. Although we might be able to use more staging nodes, that number is usually fixed (1\% of the total compute nodes the job occupies) due to limited resources. If $n,p,m,q,S$ are fixed, $t_w(),t_{r}(),t_{s}()$ are also unchangeable since they are hardware-dependent. The only variables users can control are $t_c$ and $N$. Particularly, $t_c$ determines the frequency of outputting simulation data, while $N$ determines how long the simulation runs. Therefore, we measure the performance of a given simulation setup on Summit to quantify all the unchangeable variables, and demonstrate how to select $t_c$ and $N$ to make the on-the-fly data layout reorganization more efficient than the post-hoc approach in terms of resource utilization.

We use 256 nodes and launch six processes per node to run the WarpX simulation. We also use two extra nodes and launch 32 processes per node for data staging. Each simulation output has size 256~GB, which is one 3D mesh variable (it is possible to reorganize the layout of all variables through staging, but here we only output one variable to be consistent with the read performance comparison in the previous sections
). We also test the post-hoc data layout reorganization using 256 nodes (six processes per node) for simulation and two nodes (32 processes per node) for the post-hoc run. In the reorganized layout, the 2048$\times$4096$\times$4096 mesh variable is decomposed into 64 chunks and the decomposition scheme is 4$\times$4$\times$4. Our measurements, listed in \autoref{tab:timing-info}, show that $t_{r}(2,32,256N)=t_{r}(2,32,256)N=11.1N$ while $t_{w}(2,32,256N)=t_{w}(2,32,256)N=13.6N$. 

\begin{table}[ht]
  \caption{Timing information of our staging-based data layout reorganization experiment}
  \centering
    \begin{tabular}{|c|c|}
      \hline
      \textbf{Variable} &  \textbf{Value (seconds)}\\
      \hline
      $t_{s}(n=256,p=6,m=2,q=32,S=256)$ & 19.4 \\
      \hline
      $t_{w}(m=2,q=32,S=256)$ & 13.6 \\
      \hline
      $t_{w}(n=256,p=6,S=256)$ & 1.4 \\
      \hline
      $t_{r}(m=2,q=32,S=256)$ & 11.1 \\
      \hline
    \end{tabular}
  \vspace{-2mm}
  \label{tab:timing-info}
\end{table}

If we fix $t_c=40$ seconds (i.e., the simulation outputs data every 400 steps, since each simulation takes about 0.1 seconds under this setting), then $t_{s}(n=256,p=6,m=2,q=32,S=256)+t_{w}(m=2,q=32,S=256)=19.4+13.6=33 < t_c$. The resource utilization of on-the-fly reorganization is $U_{o}=(256+2)(40N+19.4+13.6)=258(40N+33)$. The resource utilization of post-hoc reorganization is $U_{p}=256N(40+1.4)+2(11.1N+13.6N)=10647.8N$. Only if $U_{o}<U_{p}$, which means $258(40N+33)<10647.8N \Rightarrow N \geq 26$ (the simulation needs to at least outputs data 26 times), the resource utilization of on-the-fly reorganization will be less than the post-hoc reorganization.

If we fix $t_c=20$ seconds (i.e., the simulation outputs data every 200 steps), then $t_{s}(n=256,p=6,m=2,q=32,S=256)+t_{w}(m=2,q=32,S=256)=19.4+13.6=33 > t_c$. The resource utilization of on-the-fly reorganization is $U_{o}=(256+2)[20+N(19.4+13.6)]=258(33N+20)$. The resource utilization of post-hoc reorganization is $U_{p}=256N(20+1.4)+2(11.1N+13.6N)=5527.8N$. Since $258(33N+20)>5527.8N$, we always have $U_{o}>U_{p}$, meaning the resource utilization of on-the-fly reorganization will always be more than that of post-hoc reorganization if $t_{c}=20$. In fact, to make $U_{o}<U_{p}$ in this case, we need $t_c > \frac{8106.2N}{256N-258} > \lim_{N\to\infty}\frac{8106.2N}{256N-258} = 31.66$. Thus, we need to choose $t_c$ such that $31.66<t_c<33$.

If the user wants the simulation to output at least 50 outputs ($N\geq50$ ) and the simulation is not slowed down by the on-the-fly reorganization ($t_c > 33$), $t_c$ needs to be: $U_{o}<U_{p} \Rightarrow 258(Nt_c+33)<256N(t_c+1.4)+49.4N \Rightarrow t_c < \frac{407.8N-8514}{2N} \Rightarrow t_c < 150.26$, to make resource utilization of on-the-fly reorganization less than in the post-hoc approach.

\begin{figure}[ht] \centering
\includegraphics[width=\columnwidth,trim=0 0 0 7mm,clip]{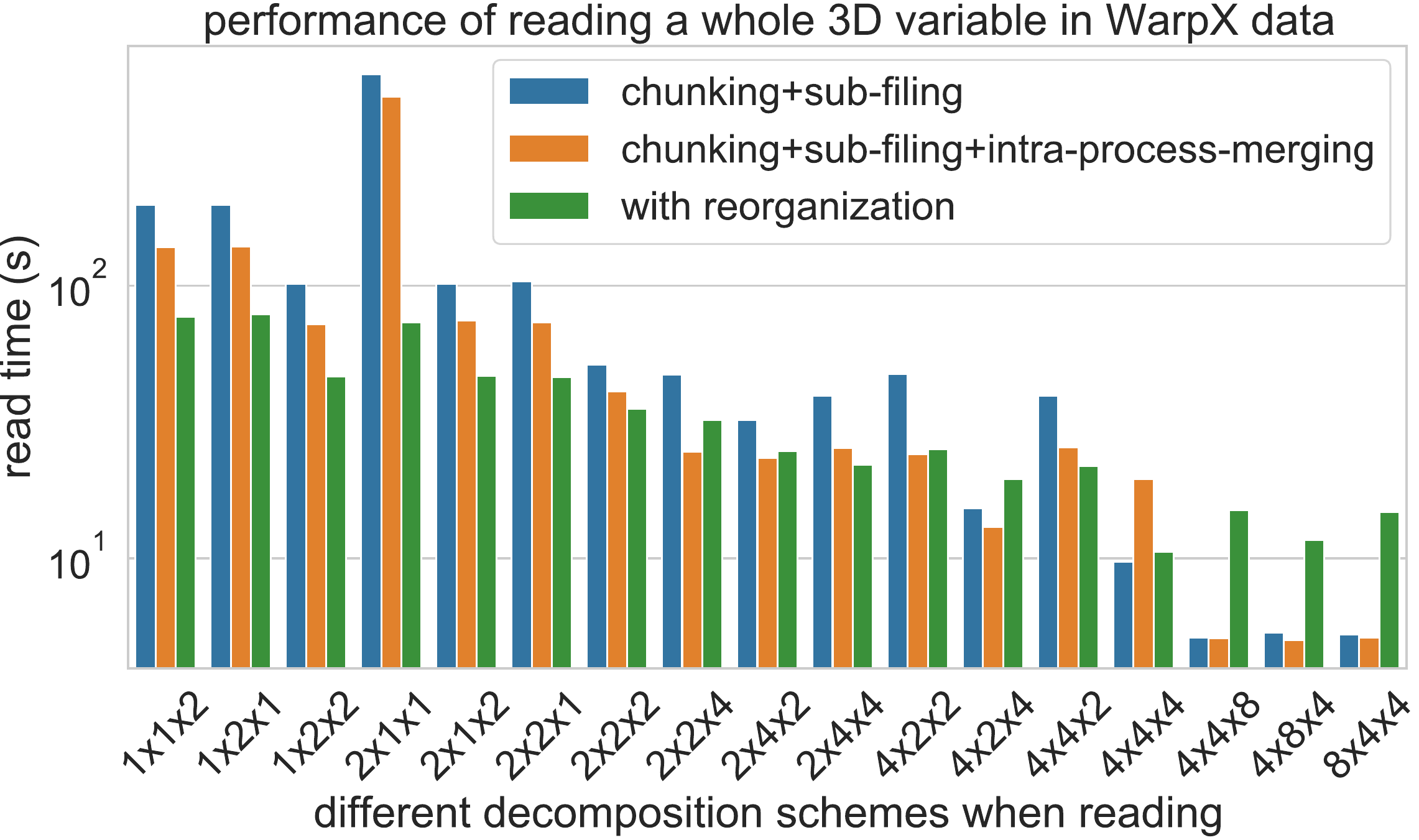}
\caption{Read performance after data layout reorganization: Reading a whole 3D variable in WarpX data}
\label{fig:with-without-reorg-read-perf}
\end{figure}

Finally, we evaluate read performance after data layout reorganization and compare it with that achieved with other approaches. As shown in \autoref{fig:with-without-reorg-read-perf}, after data layout reorganization, read performance is notably improved compared to other approaches with less than 16 concurrent readers. In particular, when reading with a 2$\times$1$\times$1 decomposition, the read time is reduced by 85\% compared to other approaches. However, as the number of concurrent readers increases, the performance improvement provided by data layout reorganization reduces; with more than 64, the read performance after data layout reorganization becomes worse than other approaches. Because the reorganized data has 64 chunks; with  more than 64 concurrent readers, one chunk might be accessed by multiple readers, leading to contention.

\section{Related Work}

Data layout is a crucial determining factor for I/O access latency and bandwidth on parallel computers. Arranging data accesses so as to increase I/O system performance, either for specific cases or as a general goal, has been studied for over two decades \cite{10.1145/244764.244766, 10.5555/358533}. Historically, parallel I/O middleware and file systems have been developed separately, to simplify implementations and enhance transparency between parallel I/O components. However, recent studies have investigated data layout-aware optimization strategies that promote a better integration of parallel I/O middleware and file systems, with promising results \cite{5600297, 10.1007/s11227-013-1077-6}. 

As the performance gap between processors and storage devices keeps increasing, more studies have focused on better understanding the IO patterns of applications and the potential benefits of pre-arranging data in different layouts. Liu et al.~\cite{6969581} investigate the impact on read performance in data analysis tasks when the read pattern does not conform with the original organization of the data. Based on the findings, the authors propose a method for automatically reorganizing previously written data to conform with the known read patterns. Tang et al.~\cite{7515710} extend this work by adding a dynamically component to the method, allowing it to recognize data usage patterns and to replicate the data of interest in multiple reorganized layouts that would benefit common read patterns at runtime.

More recent work focuses on optimizing the most currently used IO libraries in large-scale data centers for the frequent patterns within HPC applications. He et al.~\cite{8862852}focus on MPI-IO, creating methods to reorganized data replica for each access pattern on HDD-based or SSD-based servers for low or high I/O concurrency applications depending on their patterns. Similarly, Tsujita et al.~\cite{10.1145/3149457.3149464} investigate the performance benefits of aggregation methods for collective communications through the MPI-IO layer.

The performance of the HDF5 IO library has been studied extensively and multiple methods have been proposed on top of it to optimize the data access. Ji et al.~\cite{10.1007/978-3-030-38961-1_55} propose several strategies to optimize HDF5 I/O operations, including chunk storage, parallel read/write, on-demand dump, and stream processing for the Five-hundred-meter Aperture Spherical Radio Telescope project. Mehta et al.~\cite{6495884} develop a new HDF5 plugin in order to use the parallel file system to convert the single-file layout into a data layout that is optimized and stores data in a unique way that enables semantic post-processing on data. Mu et al.~\cite{mu2020interfacing}, design a storage interface based on data containers that provides data chunking and takes advantage of multiple storage tiers. 

The ADIOS library uses chunking for data access and can use the block range index technique for scientific datasets, which only records the value range of all records in a data block~\cite{7973792}. 
Lofstead et al.~\cite{10.1145/1996130.1996139} 
compare the ADIOS log-based BP format to the logically contiguous NetCDF or HDF5 formats, 
taking into account different patterns, layouts, and data sizes. Another similar study \cite{10.1145/2222} focuses on large-scale particle simulations. It 
introduces new techniques for scalable, spatially-aware write and read operations, and designs an adaptive aggregation technique to improve read operations via a multi-resolution layout.

Besides the I/O libraries, the object stores have also drawn a lot of attention from the HPC community. The object stores such as Intel's DAOS~\cite{DAOS} and Seagate's Cortex~\cite{Cortex}, 
are potential solutions for the metadata challenge exascale applications are facing.

\section{Conclusion}
We have presented a comprehensive study of common data layout strategies for parallel I/O. We show that due to the complex I/O patterns of scientific codes with dynamic load-balancing and/or AMR, no standard data layout strategies used in existing I/O libraries can achieve both satisfactory write and read performance at the same time. 

Knowing the limitations of existing data layouts, we propose two online data layout reorganization approaches, with the aim of enabling good tradeoffs between write and read performance for these complex I/O patterns. The first approach leverages spatial locality in data to cluster and merge data blocks within each process or compute node; we show that for most common read patterns, this strategy can significantly improve read performance while incurring only mimimal write overhead.
Performance results for this method, using realistically complex application I/O patterns, are readily applicable without any change to the application code.

The second approach is to fully reorganize the data layout by leveraging the existing staging techniques. 
Due to a higher degree of freedom in the reorganization, potential gains for read performance brought by this approach are even larger than in the first approach. Since large number of data blocks need to be moved among processes during the online data layout reorganization which might cause significant overhead, we build a model to understand when and how to use the staging-based data layout reorganization can achieve better resource utilization compared to the post-hoc approach.




%

\ifCLASSOPTIONcompsoc
  \section*{Acknowledgments}
\else
  \section*{Acknowledgment}
\fi

This work was supported by the Exascale Computing Project (17-SC-20-SC), a collaborative effort of the U.S.\ Department of Energy Office of Science and the National Nuclear Security Administration.
This research used resources of the Oak Ridge Leadership Computing Facility, a DOE Office of Science User Facility supported under Contract DE-AC05-00OR22725.
This work was partially funded by the Center of Advanced Systems Understanding (CASUS), financed by Germany's Federal Ministry of Education and Research (BMBF), and by the Saxon Ministry for Science, Culture and Tourism (SMWK) with tax funds on the basis of the budget approved by the Saxon State Parliament.

\ifCLASSOPTIONcaptionsoff
  \newpage
\fi



\bibliographystyle{IEEEtran}
\bibliography{related_work}
%

\vspace{-0.3in}

\vspace{-5mm}
\begin{IEEEbiography}[{\vspace{-11mm}\includegraphics[trim={5mm 3mm 5mm 0}, width=0.75in,clip,keepaspectratio]{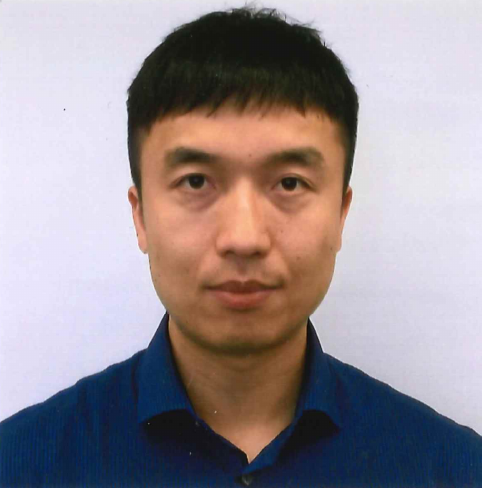}}]{Lipeng Wan}
is a Computer Scientist in the Computer Science and Mathematics Division at Oak Ridge National Laboratory (ORNL). He received his Ph.D. degree in computer science from the University of Tennessee, Knoxville in 2016. His research mainly focuses on scientific data management and high-performance computing. 
\end{IEEEbiography}
\vspace{-23mm}
\begin{IEEEbiography}[{\vspace{-9mm}\includegraphics[trim={15mm 0 15mm 0}, width=0.75in,clip,keepaspectratio]{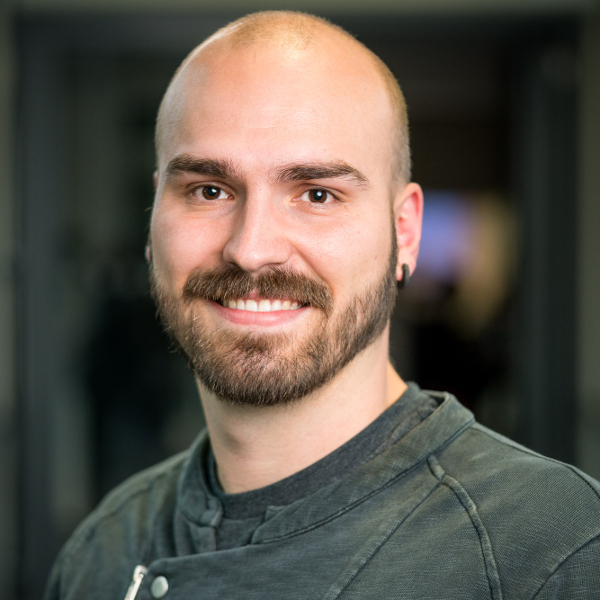}}]{Axel Huebl} (Member, IEEE)
works in the Accelerator Technology and Applied Physics (ATAP) Division at Berkeley Lab. 
He received his Ph.D. degree  in Physics from Technical University Dresden, Germany in 2019.
His research interests are in laser-plasma based particle acceleration, high-performance computing, and scalable, data-driven science.
\end{IEEEbiography}
\vspace{-20mm}
\begin{IEEEbiography}[{\vspace{-7mm}\includegraphics[trim={5mm 0 5mm 0}, width=0.75in,clip,keepaspectratio]{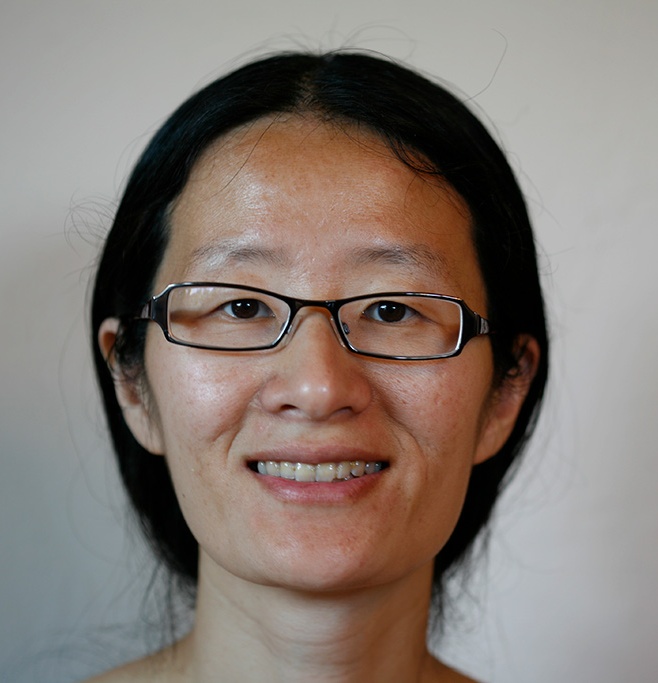}}]{Junmin Gu}
is a Computer System Engineer working in the Computational Research Division at LBNL. Her interests include data management, resource management, distributed systems and high performance computing.  She received Master degrees in Mathematics and Computer Science from the University of Wisconsin-Madison. 
\end{IEEEbiography}
\vspace{-20mm}
\begin{IEEEbiography}[{\vspace{-7mm}\includegraphics[trim={0 30mm 0 0},
width=0.75in,clip,keepaspectratio]{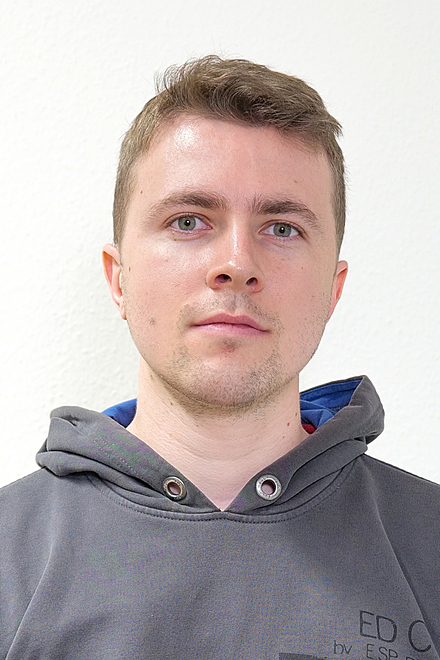}}]{Franz Poeschel} is a Computer Scientist at the CASUS Center for Advanced Systems Understanding. His interests of research include high-performance computing, large-scale IO and data staging. He is a developer and maintainer for the openPMD API. 
He received his Master of Science degree in 2020 at the Technical University of Dresden, Germany.
\end{IEEEbiography}
\vspace{-20mm}
\begin{IEEEbiography}[{\vspace{-9mm}\includegraphics[trim={140mm 90mm 60mm 170mm}, width=0.75in,clip,keepaspectratio]{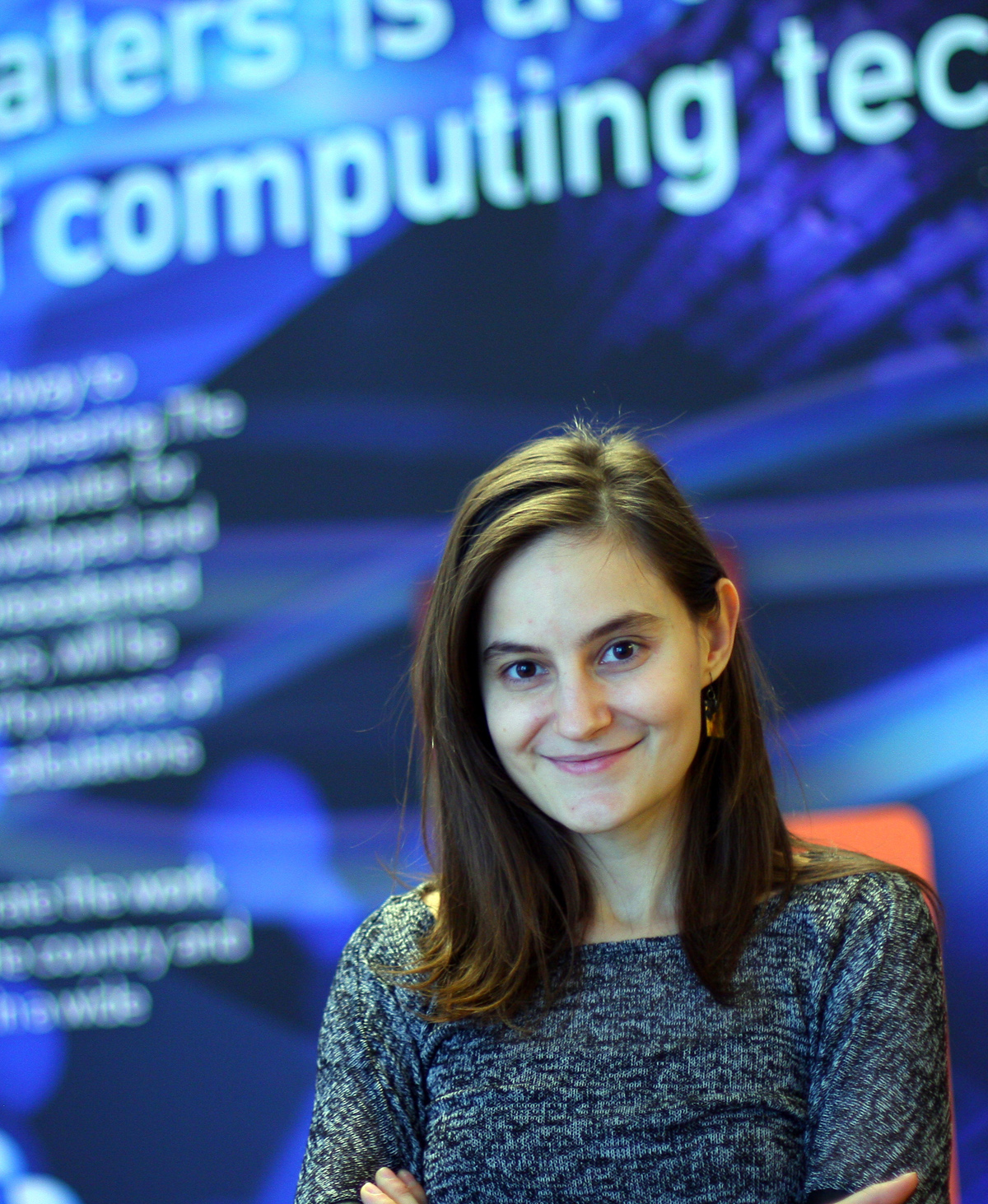}}]{Ana Gainaru} 
is a computer scientist in the CSM division at ORNL.
She did her PhD studies at the University of Illinois at Urbana-Champaign. 
She has experience in HPC working primarily on optimizing the execution of scientific applications, from data-aware runtime design to scheduling, fault tolerance and code optimization.
\end{IEEEbiography}
\vspace{-23mm}
\begin{IEEEbiography}[{\vspace{-14mm}\includegraphics[trim={150mm 130mm 130mm 40mm}, width=0.75in,clip,keepaspectratio]{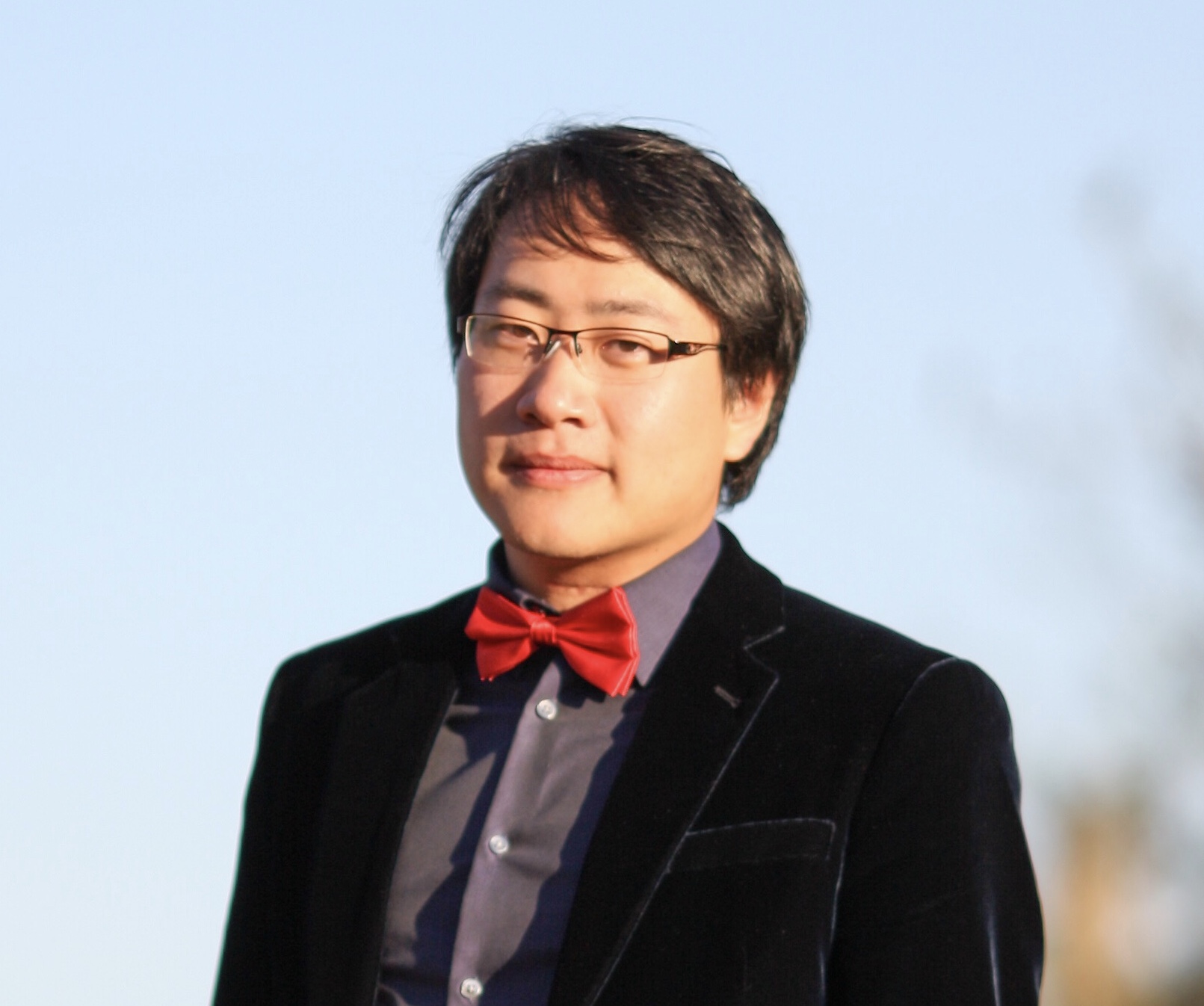}}]{Ruonan Wang} 
is a Software Engineer in the Computer Science and Mathematics Division at ORNL. He received his Ph.D. from the University of Western Australia in 2018. His work focuses on extremely large-scale I/O middleware design, and data staging techniques.   
\end{IEEEbiography}
\vspace{-24mm}
\begin{IEEEbiography}[{\vspace{-7mm}\includegraphics[trim={0 5mm 0 0}, width=0.75in,clip,keepaspectratio]{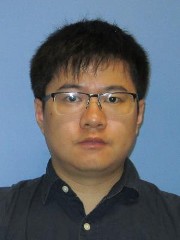}}]{Jieyang Chen} (Member, IEEE) 
is a Computer Scientist in the Computer Science and Mathematics Division at ORNL. He received his master and Ph.D. degrees in Computer Science from University of California, Riverside in 2014 and 2019. 
His research interests include high-performance computing, parallel and distributed systems, and big data analytics. 
\end{IEEEbiography}
\vspace{-18mm}
\begin{IEEEbiography}[{\vspace{-5mm}\includegraphics[trim={70mm 20mm 70mm 20mm},width=0.75in,clip,keepaspectratio]{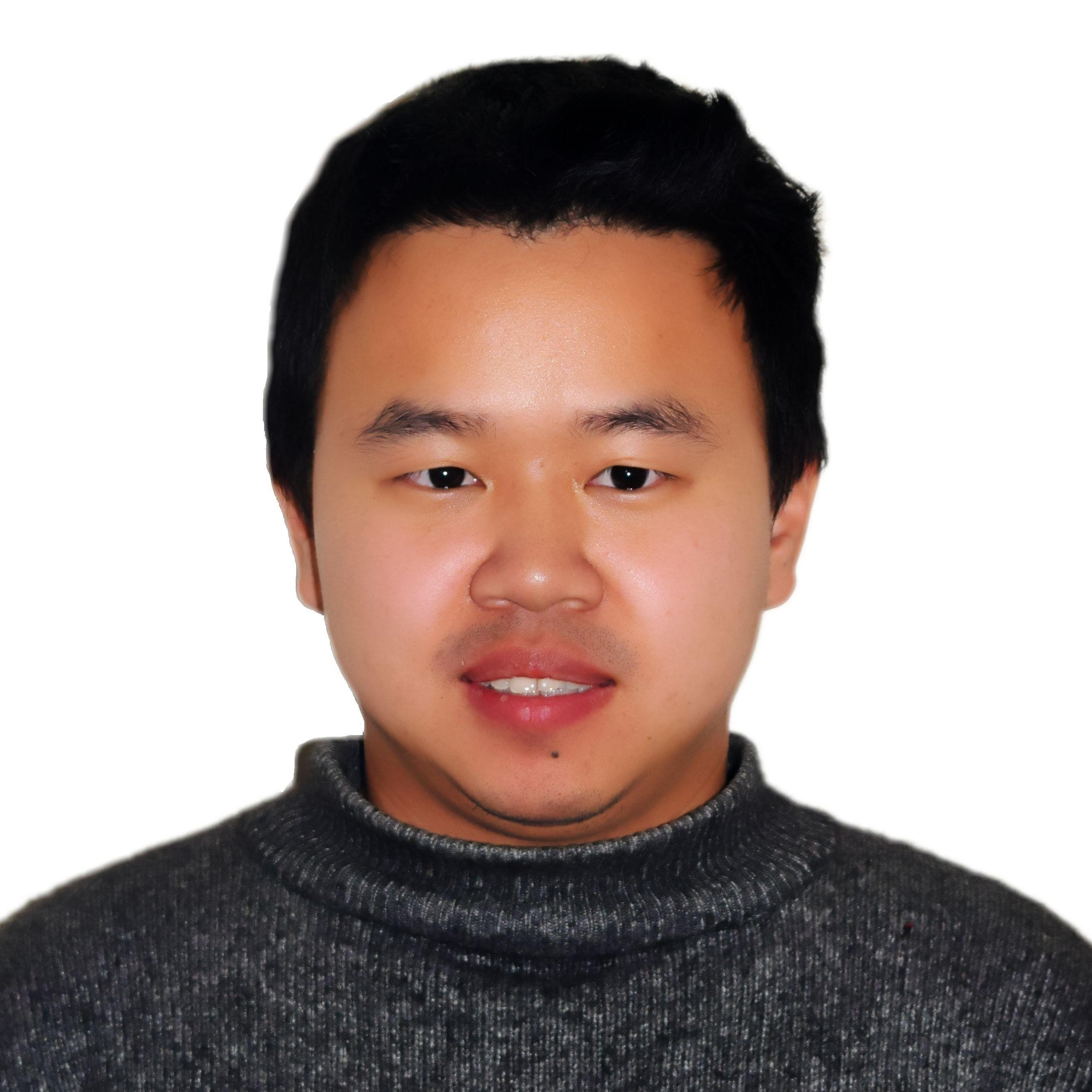}}]{Xin Liang} (Member, IEEE)
is an Assistant Professor with the Department of Computer Science at Missouri University of Science and Technology. He received his Ph.D. degree from University of California, Riverside in 2019.
His research interests include high-performance and distributed computing, 
data management and reduction, and cloud computing .
\end{IEEEbiography}
\vspace{-18mm}
\begin{IEEEbiography}[{\vspace{-7mm}\includegraphics[trim={100mm 100mm 100mm 40mm},width=0.75in,clip,keepaspectratio]{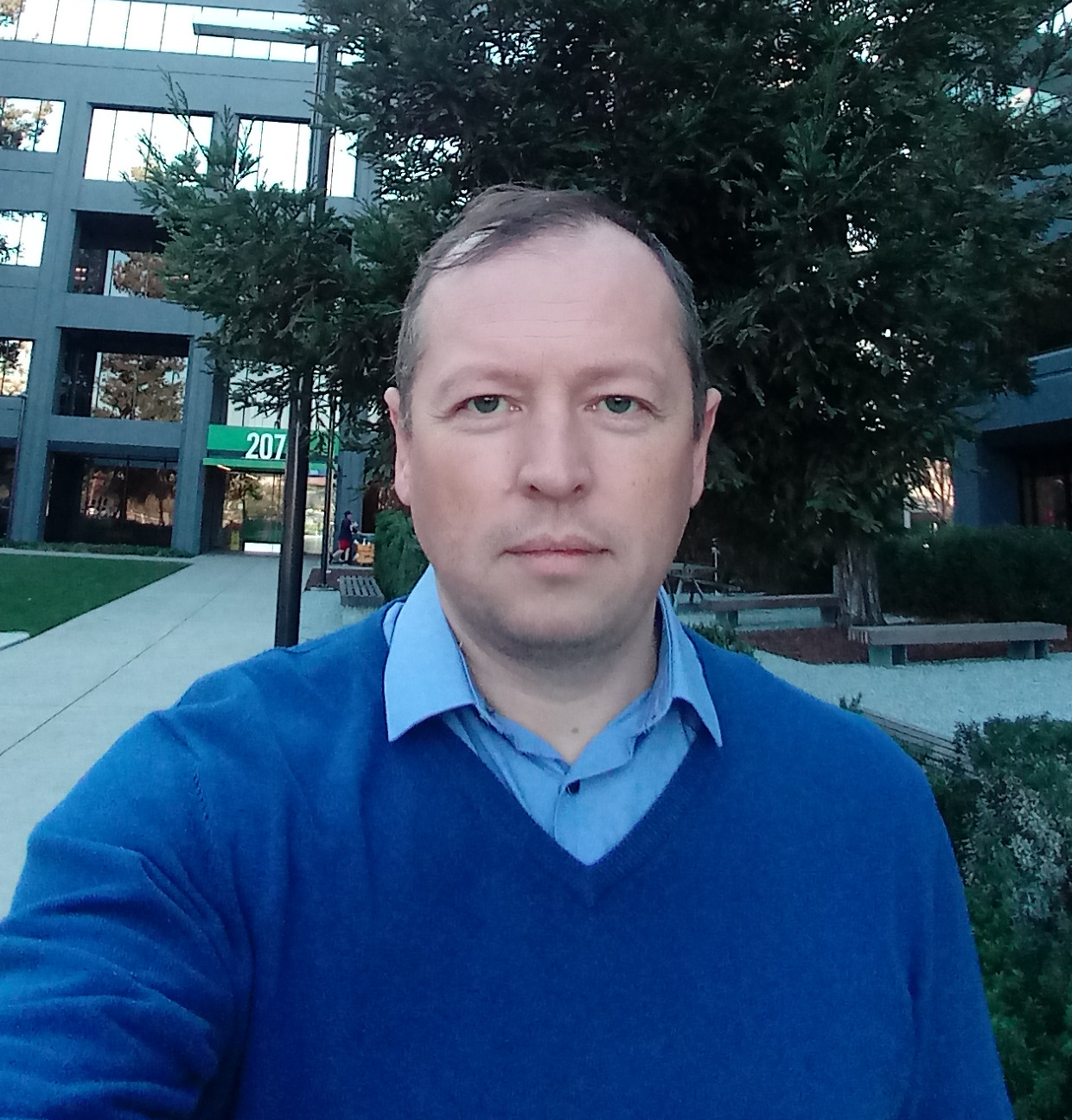}}]{Dmitry Ganyushin} 
is a Senior Software Developer in the CSM Division at ORNL. He received his Ph.D. degree in Theoretical Chemistry from Technical University of Munich, Germany in 2004. 
His research interests focus on parallel and distributed systems,  high-performance computing, and performance optimization for scientific applications. 
\end{IEEEbiography}
\vspace{-18mm}
\begin{IEEEbiography}[{\vspace{-7mm}\includegraphics[trim=25 120 40 15, width=0.75in,clip,keepaspectratio]{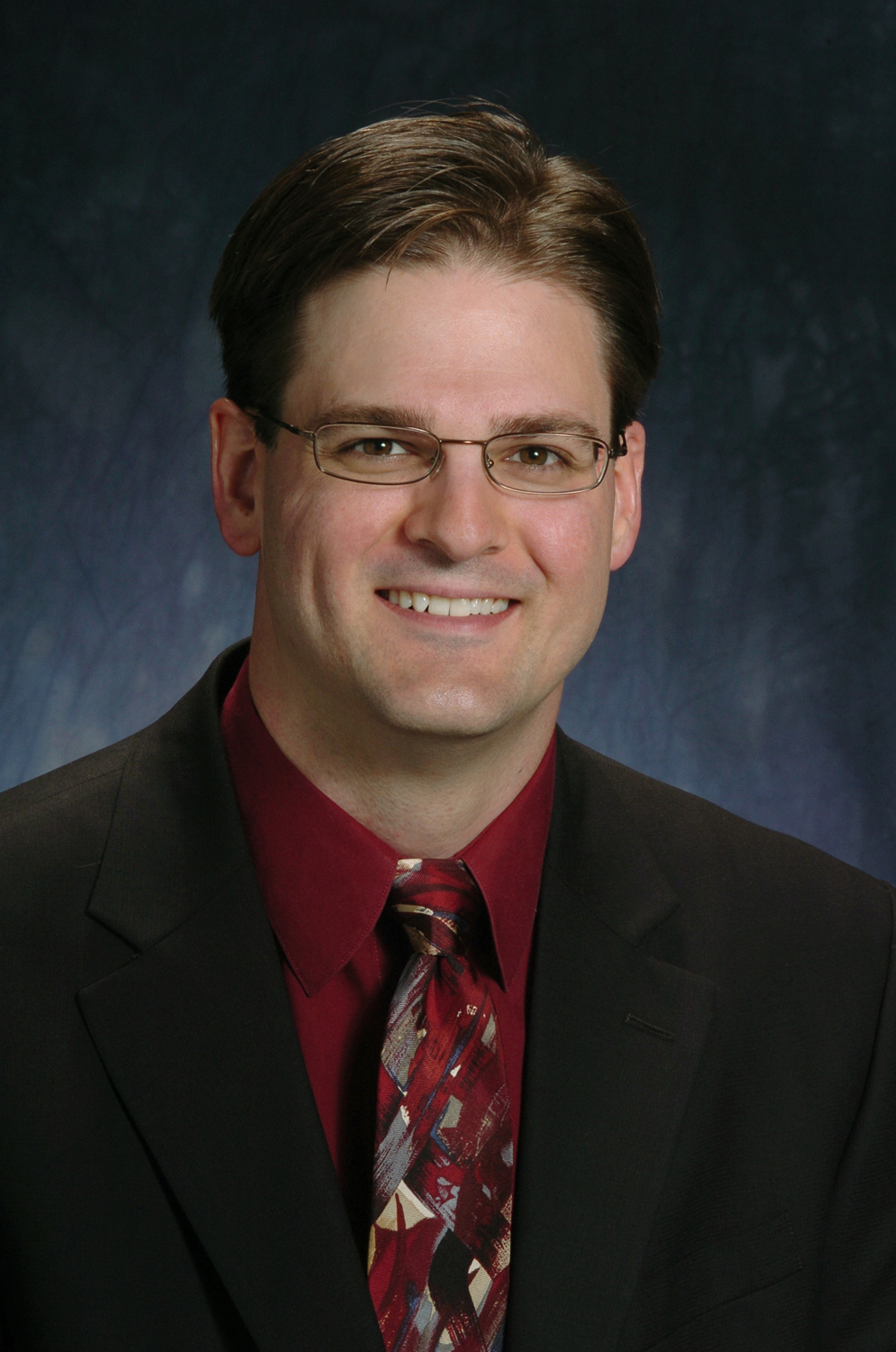}}]
{Todd Munson} is a Senior Computational Scientist
at Argonne National Laboratory
and the Software Ecosystem and Delivery Control
Account Manager for the U.S.\ DOE Exascale Computing Project.
He received his Ph.D from the University of Wisconsin at 
Madison in 2000. His interests range from numerical 
methods
to workflow optimization for online 
data analysis and reduction.
\end{IEEEbiography}
\vspace{-18mm}
\begin{IEEEbiography}[{\vspace{-7mm}\includegraphics[trim={2mm 0 2mm 0},
width=0.75in,clip,keepaspectratio]{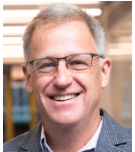}}]
{Ian Foster} (Fellow, IEEE) is a Senior Scientist, Distinguished Fellow, and director of the Data Science and Learning Division, at Argonne National Laboratory, and the Arthur Holly Compton Distinguished Service Professor of Computer Science at the University of Chicago. His research deals with distributed, parallel, and data-intensive computing technologies.
\end{IEEEbiography}
\vspace{-18mm}
\begin{IEEEbiography}[{\vspace{-8mm}\includegraphics[trim={5mm 0 5mm 0}, width=0.75in,clip,keepaspectratio]{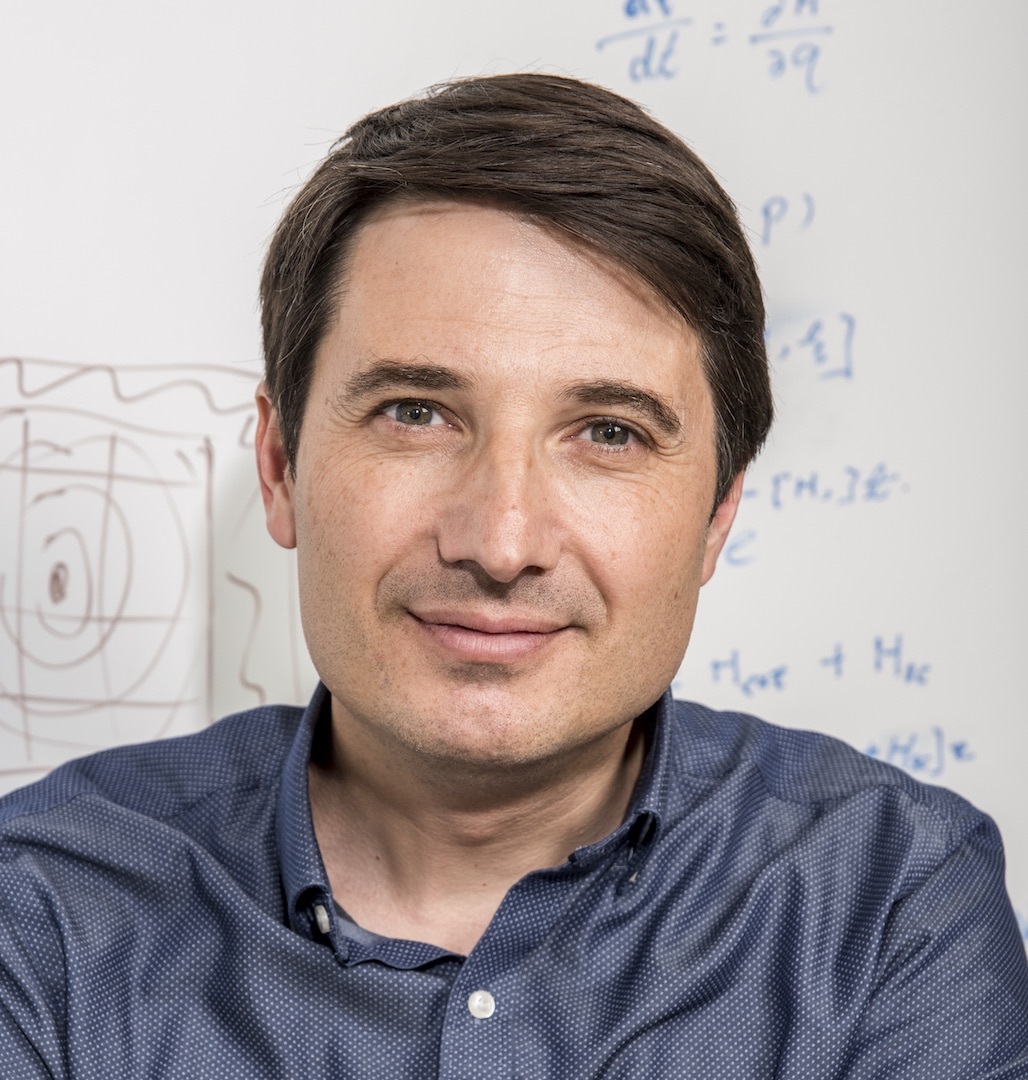}}]{Jean-Luc Vay} (Senior Member, IEEE) heads the Accelerator Modeling Program in the ATAP Divison at Berkeley Lab. He received his Ph.D. in Physics from the University of Paris, France. His interests include the development of algorithms, their optimization on supercomputers, and their application to particle accelerator modeling.
\end{IEEEbiography}
\vspace{-18mm}
\begin{IEEEbiography}[{\vspace{-7mm}\includegraphics[trim={5mm 15mm 5mm 5mm}, width=0.75in,clip,keepaspectratio]{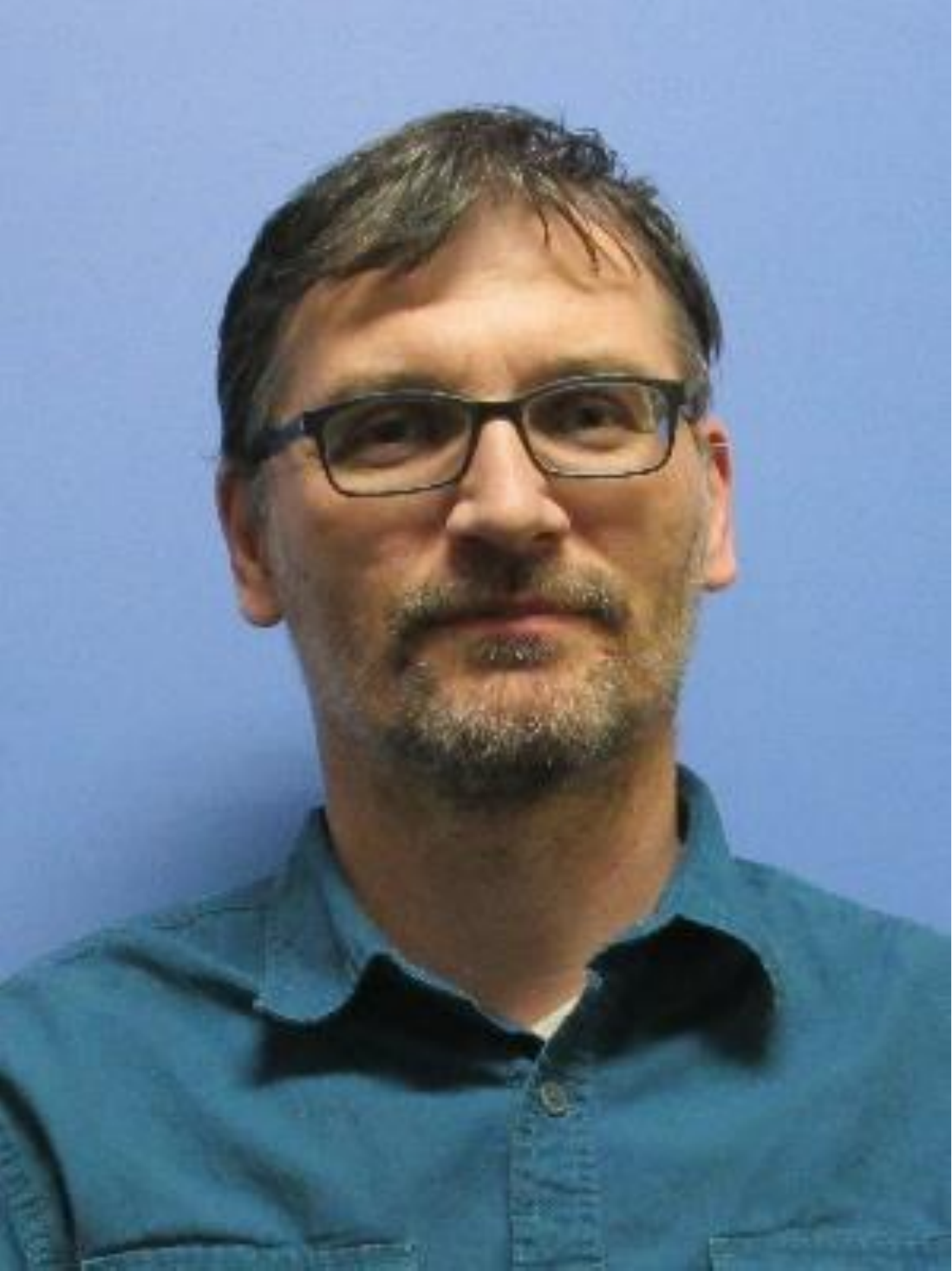}}]{Norbert Podhorszki}
Norbert Podhorszki is a Senior Research Scientist at ORNL in the CSM division. He is one of the key developers of ADIOS.
His main research interest is in creating I/O and staging solutions for in-situ processing of data on leadership class computing systems. He received his Ph.D. in Information Technology from the Eötvös Loránd University of Budapest. 
\end{IEEEbiography}
\vspace{-18mm}
\begin{IEEEbiography}[{\vspace{-5mm}\includegraphics[trim={2mm 0 2mm 0}, width=0.75in,clip,keepaspectratio]{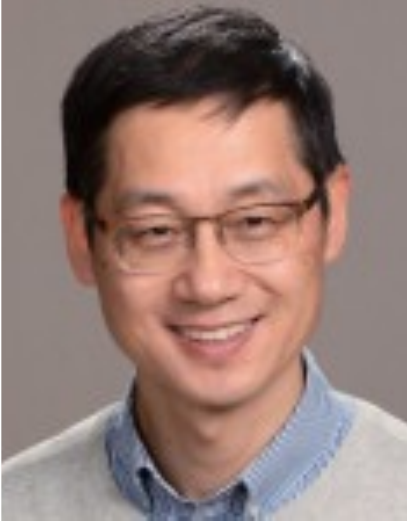}}]{Kesheng Wu} 
is a Senior Computer Scientist at LBL. He received a Ph.D. in computer science from the University of Minnesota. His research primarily focuses on improving bitmap index technology with compression, encoding and binning. He is the key developer of FastBit bitmap indexing software, which has been used in a number of applications.
\end{IEEEbiography}
\vspace{-18mm}
\begin{IEEEbiography}[{\vspace{-5mm}\includegraphics[trim={2mm 0 2mm 0}, width=0.75in,clip,keepaspectratio]{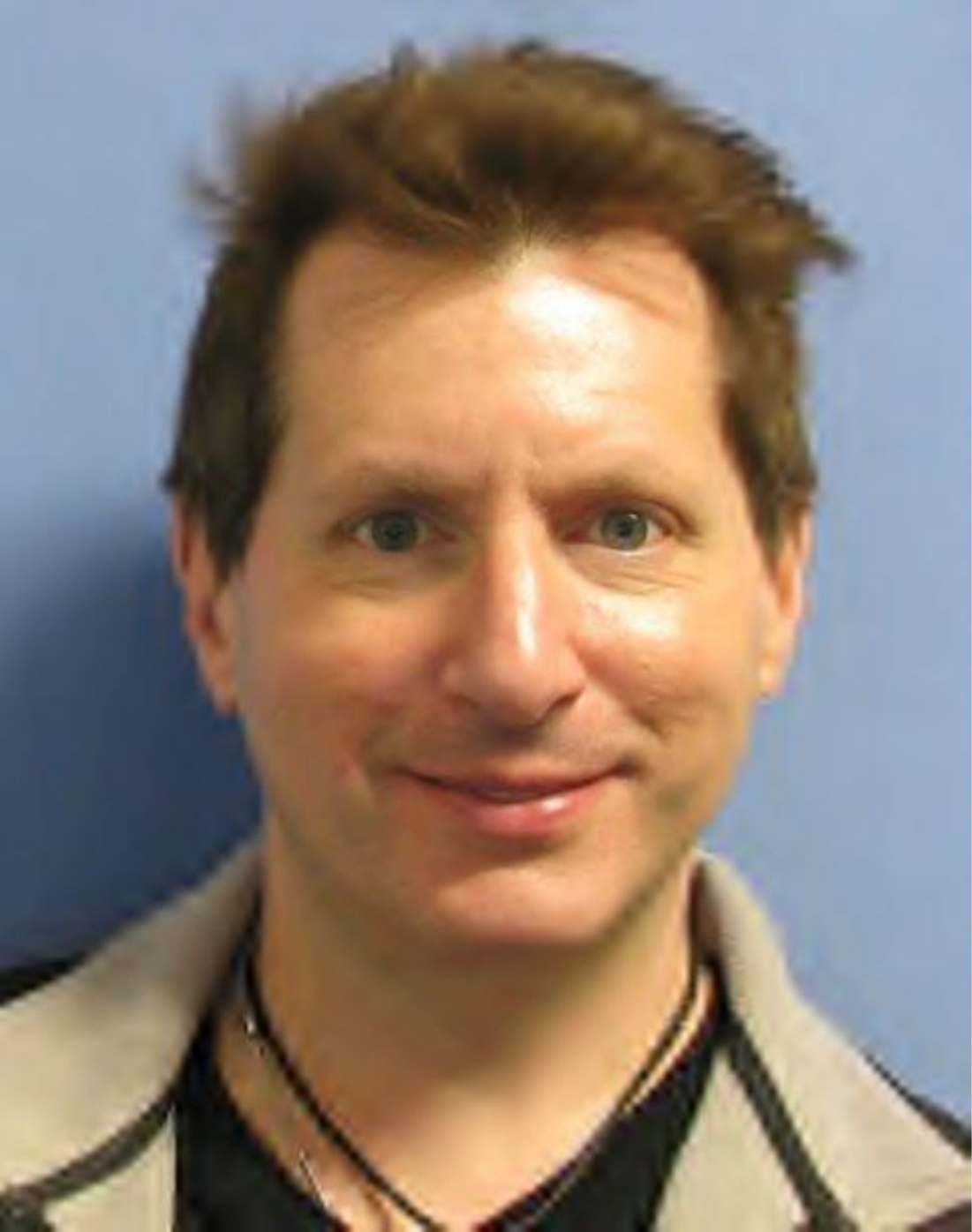}}]{Scott Klasky} (Senior Member, IEEE)
is a distinguished scientist and Group Leader in the CSM Division at ORNL. He also
has a joint faculty appointment at the University of Tennessee, Knoxville, and an adjunct position at Georgia Tech.  
He received his Ph.D. in Physics from the University
of Texas at Austin. 
He has expertise in HPC, data management, workflow automation, data reduction, visualization and physics. 
\end{IEEEbiography}

\end{document}